\documentclass[useAMS,usenatbib]{mn2e}
\usepackage{graphicx}

\title[The Octave pipeline for extracting oscillation parameters ]{The Octave (Birmingham - Sheffield Hallam) automated pipeline for extracting oscillation parameters of solar-like main-sequence stars}
\author[S. Hekker et al. 2009]{S. Hekker$^{1}$\thanks{E-mail: saskia@bison.ph.bham.ac.uk}, A.-M Broomhall$^{1}$, W.~J. Chaplin$^{1}$, Y.~P. Elsworth$^{1}$, S.~T. Fletcher$^{2}$,
\newauthor  R. New$^{2}$, T. Arentoft$^{3}$, P.-O. Quirion$^{3}$, H. Kjeldsen$^{3}$\\
$^{1}$University of Birmingham, School of Physics and Astronomy, Edgbaston, Birmingham B15 2TT, UK\\
$^{2}$ Materials Engineering Research Institute, Faculty of Arts, Computing, Engineering and Science, Sheffield Hallam University,\\ Howard Street, Sheffield S1 1WB, UK\\
$^{3}$ Department of Physics and Astronomy, Aarhus University, DK-8000 Aarhus C, Denmark}

\begin{document}

\date{Accepted: Retreived: }

\pagerange{\pageref{firstpage}--\pageref{lastpage}} \pubyear{2009}

\maketitle

\label{firstpage}

\begin{abstract}
The number of main-sequence stars for which we can observe solar-like oscillations is expected to increase considerably with the short-cadence high-precision photometric observations from the NASA Kepler satellite. Because of this increase in number of stars, automated tools are needed to analyse these data in a reasonable amount of time. In the framework of the asteroFLAG consortium, we present an automated pipeline which extracts frequencies and other parameters of solar-like oscillations in main-sequence and subgiant stars. The pipeline uses only the timeseries data as input and does not require any other input information. Tests on 353 artificial stars reveal that we can obtain accurate frequencies and oscillation parameters for about three quarters of the stars. We conclude that our methods are well suited for the analysis of main-sequence stars, which show mainly p-mode oscillations. 
\end{abstract}

\begin{keywords}
methods: data analysis -- stars: main-sequence -- stars: oscillations
\end{keywords}

\section{Introduction}
Stars with sub-surface convection zones, like the Sun, display acoustic oscillations. The stochastic excitation mechanism limits the amplitudes of the oscillations to intrinsically weak values. However, it gives rise to a rich spectrum of oscillations. The excited pressure (p) modes probe different interior volumes, with the radial and other low angular-degree modes probing as deeply as the core. This differential penetration of the oscillations allows the internal structure and dynamics to be inferred as a function of depth. Seismic studies of the Sun have indeed proven to be very powerful in inferring its internal structure.

The fact that the solar-like oscillations have such small amplitudes has made observations of these oscillations in stars other than the Sun very challenging. Over the past few years asteroseismic observations of main-sequence stars up to evolved red-giant stars have been made using Doppler velocity measurements from ground-based spectrographs, e.g. ELODIE \citep{baranne1996}, CORALIE, HARPS \citep{queloz2001}, UCLES and UVES \citep{odorico2000}, and photometric space-based instruments such as WIRE \citep{buzasi2000}, MOST \citep{matthews2000} and CoRoT \citep{baglin2006}. This has led to detections of solar-like oscillations in more than ten main-sequence stars and of the order of one thousand red giants. For a recent, but pre-CoRoT, review of these results see \citet{bedding2008}. CoRoT results for main-sequence stars are presented by e.g., \citet{michel2008,appourchaux2008,garcia2009}, while \citet{deridder2009} and \citet{hekker2009} present first CoRoT results for red giants.

The NASA Kepler satellite was launched successfully into an Earth trailing orbit on March 7, 2009. The satellite contains a Schmidt telescope with a 0.95-m aperture and a 105-deg$^2$ field of view, equipped with a highly sensitive photometer with a spectral bandpass from 400 to 850~nm. It is designed to continuously and simultaneously monitor 100\,000 stars brighter than 14$^{\rm th}$ magnitude. Kepler will be pointed towards the constellations Cygnus and Lyra during the entire mission, which has a nominal length of 3.5 years. For most stars, data will be integrated over 30 minutes, while for approximately 512 stars at a time, data with a 1-minute cadence will be obtained. Although the driving goal for the development of Kepler is to observe transiting Earth-like exo-planets, these observations are very well suited for asteroseismology, including low-amplitude solar-like oscillations. Even so, asteroseismology will contribute to the exo-planet investigations by determining radii of the planet-hosting stars, which are needed to extract the planetary radii. The radii of stars exhibiting solar-like oscillations can be obtained using the difference in frequency between modes with consecutive radial orders, i.e., the large separation ($\Delta \nu$). This is a measure of the sound travel time across the star, which depends on the density of the star. See \citet{stello2009} for an overview of current approaches to obtain radii. The asteroseismic potential of Kepler is described in more detail by \citet{jcd2008c}.

Apart from the determination of radii, the detection of solar-like oscillations in stars at different epochs along stellar evolutionary life cycles offers the prospect to test theories of stellar evolution and stellar dynamos for many stars. The input data for probing stellar interiors are the mode parameters. Accurate mode parameters are a vital prerequisite for robust, accurate inference on the fundamental stellar parameters.

We expect in total of the order of a thousand solar-like stars to be observed by Kepler in short cadence. Short-cadence oscillation data are needed to observe solar-like oscillations in main-sequence stars and subgiants as these occur at frequencies of the order of a few-hundred up to several-thousand micro-Hertz. The short-cadence (1 minute) data have a Nyquist frequency of $\sim$ 8300 $\mu$Hz, while the Nyquist frequency of the long-cadence (30 minute) data is $\sim$ 275 $\mu$Hz. 

In preparation for the Kepler mission, the asteroFLAG consortium has developed automated tools to analyse solar-like oscillations in main-sequence stars \citep[e.g.][]{huber2009,mathur2009,mosser2009}, and tested these tools extensively, see e.g., \citet{chaplin2008,stello2009}. Automated tools are needed to cope with the large number of stars we expect to be observed with Kepler. Here, we present an automated pipeline, built to determine oscillation parameters of solar-like oscillations in main-sequence stars and subgiants, which we describe in Section 3 (some mathematical details are deferred to Appendix A). We compare the results (Section 4) of the automated analysis with the input parameters used for simulations of realistic artificial data of a few hundred stars, prepared as if they were observed with Kepler in short-cadence (Section 2). 

\section{Simulated data}
The simulated time-series are based on stellar parameters available in the Kepler Input Catalogue (KIC) \citep{brown2005} for the main-sequence and subgiants commissioning and survey targets. For these simulations, stellar parameters are randomly chosen within the expected formal and systematic errors around their KIC values and used as inputs to the model grid prepared for the Aarhus Kepler pipeline (Quirion et al. (2009), in preparation). The resulting parameters and model frequencies have then been used for the simulations. Rotation effects, granulation, activity and white noise corresponding to the brightness of the target have been added. Also the lifetimes of the oscillation modes have been varied. All time series are 30 days long with a 1-minute cadence.

The stellar models were generated with the Aarhus stellar evolution code \citep{jcd2008b} using the OPAL equation of state \citep{iglesias1996} along with the \citet{grevesse1993} solar mixture using the OPAL and \citet{alexander1994} opacity tables. The frequencies of the p modes were calculated using the adiabatic pulsation code ADIPLS \citep{jcd2008a}. The timeseries are generated using a combination of the asteroFLAG and Aarhus simulators \citep{chaplin2008,stello2004}.

\section{Methodology}
We developed an automated pipeline to obtain the following (oscillation) parameters of main-sequence and subgiant solar-like oscillators from Fourier spectra of the time-series observations:
\begin{itemize}
\item frequency range of oscillations,
\item frequency at which maximum oscillation power occurs,
\item parameterisation of the background of the entire Fourier spectrum,
\item average large frequency separation between consecutive radial orders,
\item maximum mode amplitude and amplitude envelope of the oscillations,
\item linewidth (lifetime) at the frequency of maximum power of the oscillations,
\item individual frequencies.
\end{itemize}
Our methods to determine these parameters are described below, with some mathematical details and error calculations in Appendix A. 

We stress here that this pipeline only uses the time-series data as input and does not require any other information.

\subsection{Frequency range of the oscillations} We are looking for high-order, low-degree, solar-like (p-mode) oscillations, the frequency ($\nu_{n,\ell}$) of which we expect to follow approximately the asymptotic relation \citep{tassoul1980}. In the present study we use the following version of the relation:
\begin{equation}
\nu_{n,\ell}=\Delta \nu (n+\frac{1}{2}\ell+\epsilon)-\ell(\ell+1)D,
\label{asymptotic}
\end{equation}
where $n$ is the radial order and $\ell$ the angular degree. $\Delta \nu$ is the large separation, which is sensitive to the sound travel time across the star and $\epsilon$ is a constant sensitive to the surface layers. $D$ is related to the small separation ($\delta \nu_{02}$) between adjacent modes $\ell$~=~0 and $\ell$~=~2 by the expression $\delta \nu_{02}$ $\approx$ 6$D$. Because we deal with photometric data it is unlikely that we can observe $\ell$~=~3 modes due to cancellation effects. 

We know from theoretical models and observations in the Sun and other stars that $\Delta \nu$, $\delta \nu_{02}$ and $\epsilon$ depend slightly on frequency and angular degree. Because the changes in $\Delta \nu$ are usually relatively small, we consider $\Delta \nu$ constant to a first approximation, and search for a frequency range in which the power spectrum has peaks at near-equidistant intervals. 

From 200 $\mu$Hz up to the Nyquist frequency, the power spectrum is divided into windows of variable width ($w$) depending on the location of the central frequency of the window ($\nu_{\rm central}$), with $\nu_{\rm central}$ separated by $w$/4. The frequency $\nu_{\rm central}$ is used as a proxy of $\nu_{\rm max}$ and is therefore expected to scale with the acoustic cut-off frequency. Hence, $w$ is defined as $w$~=~($\nu_{\rm central}$/$\nu_{\rm max \odot}$)~$\cdot$~$w_{\odot}$, with $\nu_{\rm max \odot}$ = 3100 $\mu$Hz, i.e., the central frequency of the oscillations in the Sun, and $w_{\odot}$ = 2000 $\mu$Hz, the expected width of the frequency interval over which we may find oscillations in the Sun, were the Sun to be observed as a bright star with Kepler.

\begin{figure}
\begin{minipage}{\linewidth}
\centering
\includegraphics[width=\linewidth]{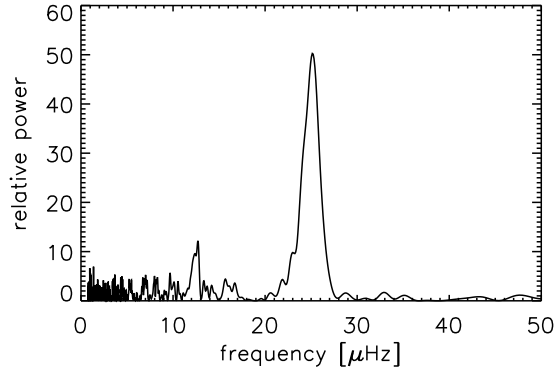}
\end{minipage}
\caption{Power spectrum of the power spectrum in a window with equidistant frequency peaks. The power is normalised, such that the noise level is 1. The $\Delta \nu$/2 at $\sim$25 $\mu$Hz and $\Delta \nu$/4 at $\sim$12 $\mu$Hz features are clearly visible.}
\label{PSPS}
\end{figure}

To find equidistant frequency peaks, we compute the power spectrum of the power spectrum (PS$\otimes$PS), which is equivalent to the autocorrelation of the time series, in each frequency window (see Fig.~\ref{PSPS} for an example of a PS$\odot$PS). Subsequently, we check for the presence of features at predicted values of $\Delta \nu$/2, $\Delta \nu$/4 and $\Delta \nu$/6. The predicted value of $\Delta \nu$ is obtained from $\Delta \nu$ $\sim$ $\nu_{\rm central}^{0.77}$ (see \citet{stello2009a} and \citet{hekker2009}), and we allowed for a 30\% deviation from the predicted value. When the probability of the presence of these three features being due to noise is less than 0.2$\%$, we interpret this as oscillations being present in the considered window. All windows in which we find equidistant frequency peaks are selected as part of the frequency range of the oscillations. For details on the computation of the probability see Appendix A.

In subgiants, for which we expect oscillations in the frequency range 100 - 1000 $\mu$Hz, the assumption of regularly spaced frequencies may no longer be valid due to the presence of g-modes and mixed modes. The increased luminosity of these more extended stars results in higher mode amplitudes ($A$). This is because $A$~$\sim$~($L$/$M$)$^s$, with $L$ and $M$ the stellar luminosity and mass, respectively. The value of exponent $s$ is of the order of 1, but still debated in the literature, e.g., see  \citet{kjeldsen1995} and \citet{samadi2005}. As a result of the increased mode amplitudes we expect oscillations with a good signal-to-noise ratio and therefore the presence of prominent peaks in the power spectrum. Therefore, in case we did not find equidistant frequency peaks, we fit a background signal including granulation, activity and white noise to the power spectrum and check whether there is a significant power excess with respect to this fit in the frequency range 100 - 1000 $\mu$Hz (see Fig.~\ref{bgfits} for examples of such fits). If this is the case, the frequency range of this power excess is taken to be the oscillation frequency range. We note here that before applying this fitting to real data one has to check for possible observational artefacts in the data, which can possibly have a similar signal in the power spectrum.

In case we do not detect any interval with equidistantly spaced frequencies, or significant power excess due to oscillations, we search once more for oscillation frequencies separated by $\Delta \nu$, but now we do not assume $\Delta \nu$ to be constant with frequency. To account for this frequency dependency, we stretch (or compress) the frequency axis of the power spectrum slightly. 
This stretching is performed in such a way as to produce an equidistant pattern of peaks on the stretched, as opposed to the original, frequency axis. The PS$\otimes$PS of the stretched power spectrum will therefore show a stronger (more prominent) signature of the large spacing than the PS$\otimes$PS of the original spectrum. 

The stretching depends on the value of $\Delta \nu$ and on the frequency range of the oscillations. We assume a maximum of 10$\%$ change in $\Delta \nu$ over the oscillation frequency range. The maximum stretch ($s_{\rm max}$) is therefore:
\begin{equation}
s_{\rm max}=0.1\Delta \nu \cdot \frac{\rm frequency ~ range}{\Delta \nu}.
\label{maxstretch}
\end{equation}
Then we compute the stretched frequency ($\nu_{\rm stretch}$) as follows:
\begin{equation}
\nu_{\rm stretch}= (\nu-\nu_{\rm c}) - j \cdot s_{\rm max} \cdot \left( \frac{\nu}{\nu_{\rm c}} -1 \right)^2,
\label{stretch}
\end{equation}
where $\nu_{\rm c}$ denotes the central frequency of the considered frequency range, which will usually be the frequency at which maximum power occurs, i.e., $\nu_{\rm max}$. Furthermore, $j$ is an integer which may have both positive and negative values, i.e. negative stretching means effectively compressing the power spectrum. To find the optimum stretch value we search for the value of $j$ for which we find minimum probability of the features in the PS$\otimes$PS to be due to noise. 


\subsection{Background signal} A background signal ($bg$) consisting of granulation, activity and white noise is fitted to a binned power spectrum where we computed the average power over independent bins. The frequency range of the oscillations is excluded. Granulation and activity are represented by power laws, from which we obtain the time scales ($\tau_{\rm gran}$ and $\tau_{\rm act}$) and power ($p_{\rm gran}$ and $p_{\rm act}$) of both phenomena respectively (see Eq.~\ref{backgroundfit}). The granulation exponent $a$ is left as a free parameter, while the activity exponent is fixed to 2. Fixing the activity exponent is justified by the fact that we assume an exponential decay of the activity over time. For the granulation the exponent is a free parameter, as in the original Harvey model \citep{harvey1985} the granulation is modelled with three exponentially decaying power laws. For the present data, we can only fit one power law for the granulation due to the limited resolution and the input in the simulations, and therefore we do not assume exponential decay, i.e., fix the exponent at 2.  Note that the background in the simulated data on which we tested our pipeline was comprised of two power laws. Should we find that more than two power laws are needed in real data, it will be straight forward to modify the code accordingly. In addition to the power laws, we add an offset $b$ which contains mostly white noise. However, in cases where no oscillations can be detected some oscillation signal might also be present in this offset. The final form of the background model used is:
\begin{equation}
bg = \frac{p_{\rm gran}}{(1+(\tau_{\rm gran} \cdot \nu)^a)} + \frac{p_{\rm act}}{(1+(\tau_{\rm act} \cdot \nu)^2)} + b.
\label{backgroundfit}
\end{equation}

\begin{figure}
\begin{minipage}{\linewidth}
\centering
\includegraphics[width=\linewidth]{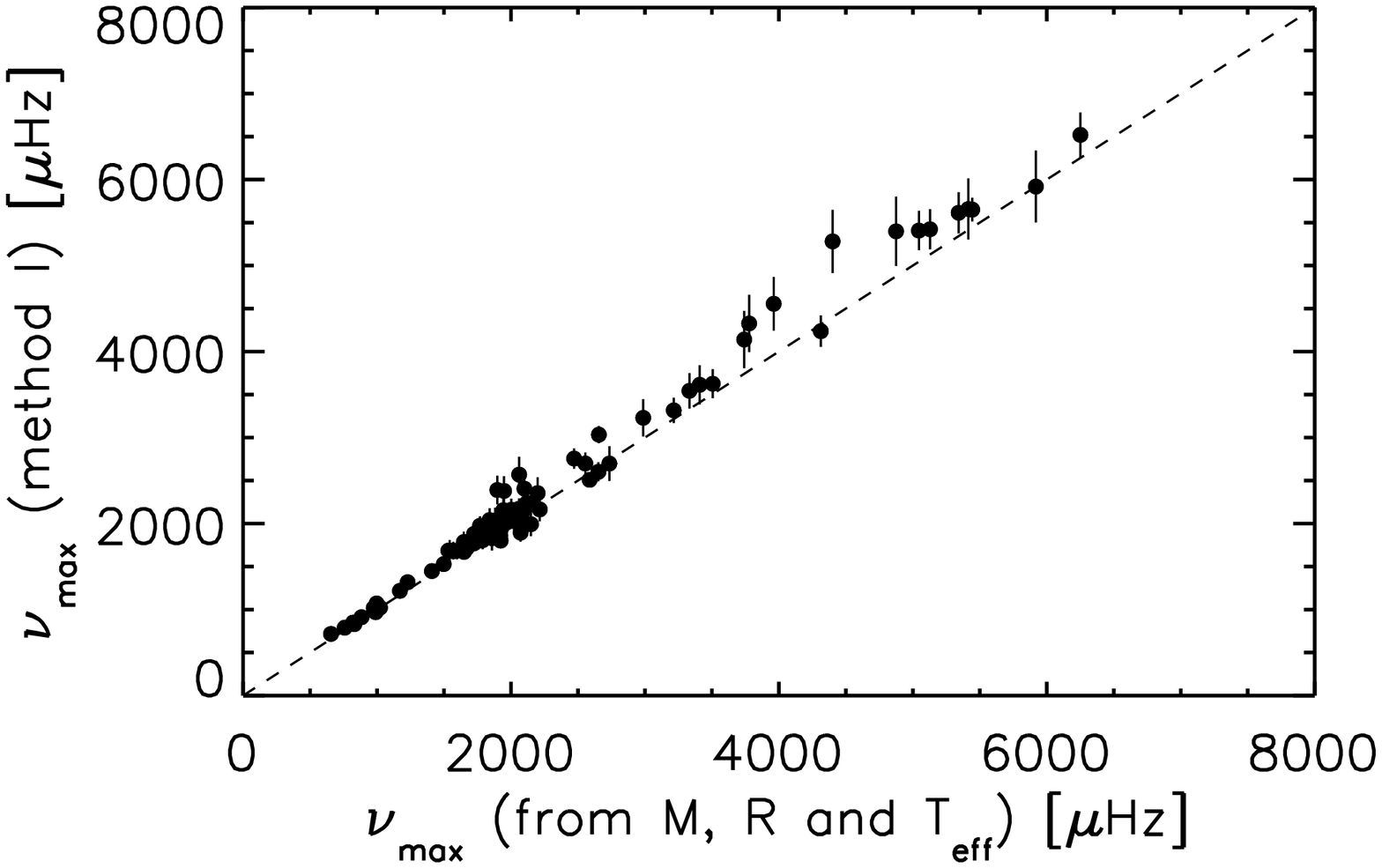}
\end{minipage}
\begin{minipage}{\linewidth}
\centering
\includegraphics[width=\linewidth]{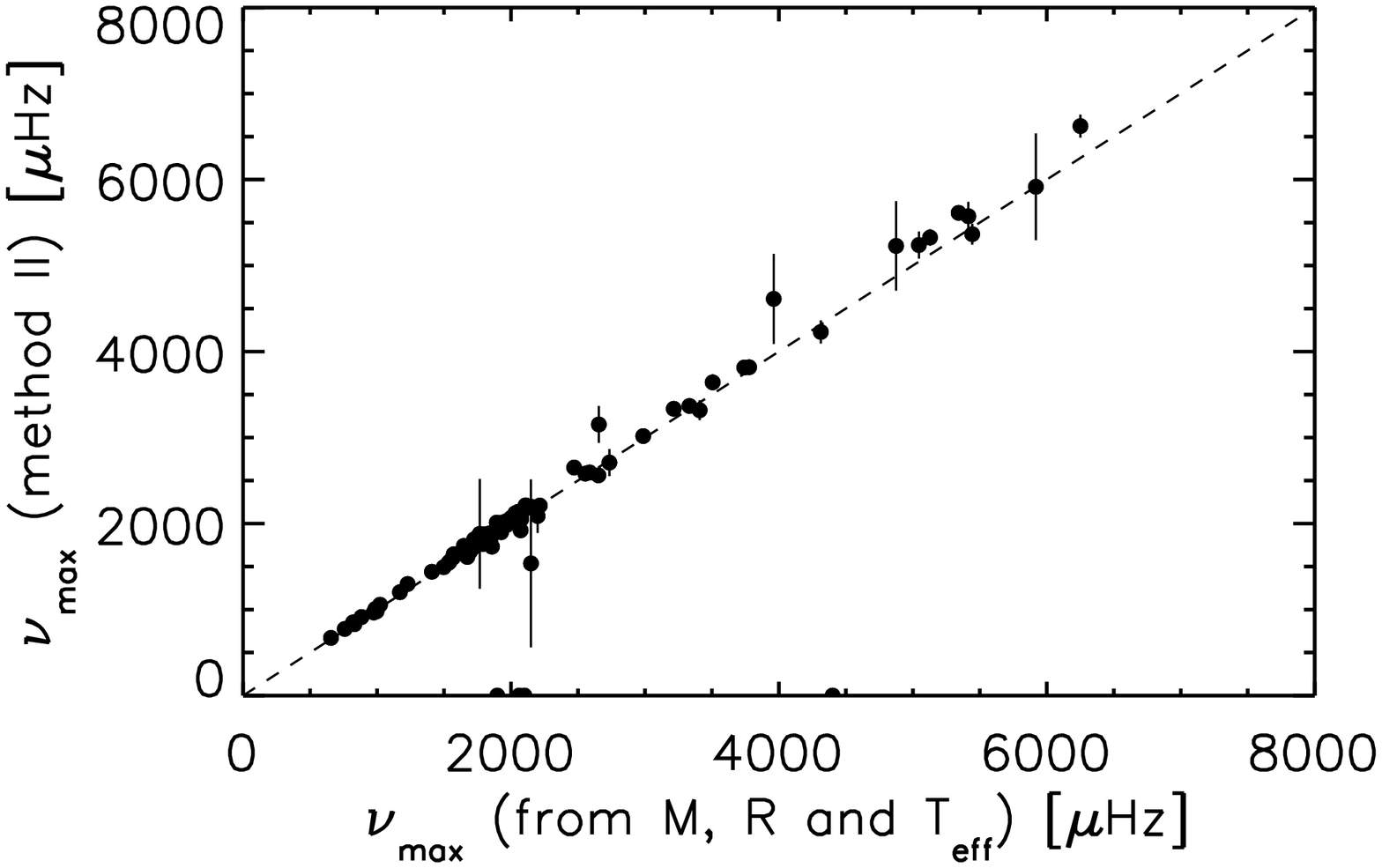}
\end{minipage}
\caption{$\nu_{\rm max}$ results from method I (top) and method II (bottom) (see Section 3.4) vs. $\nu_{\rm max}$ computed using the input $M$, $R$ and $T_{\rm eff}$ in Eq.~\ref{scale1}. The one-to-one relations are indicated with a dashed line.}
\label{numax}
\end{figure}

\begin{figure*}
\begin{minipage}{0.45\linewidth}
\centering
\includegraphics[width=\linewidth]{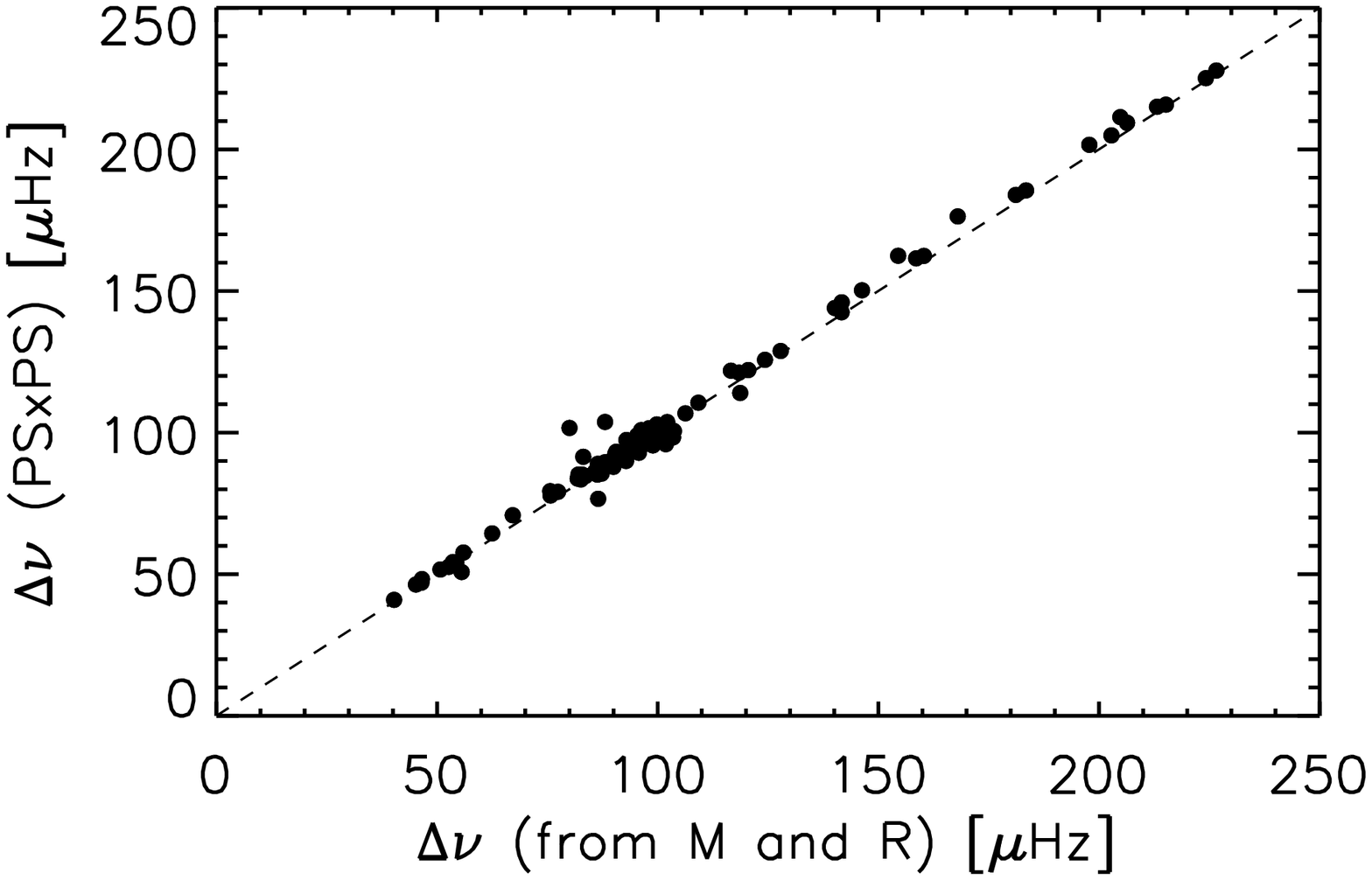}
\end{minipage}
\begin{minipage}{0.45\linewidth}
\centering
\includegraphics[width=\linewidth]{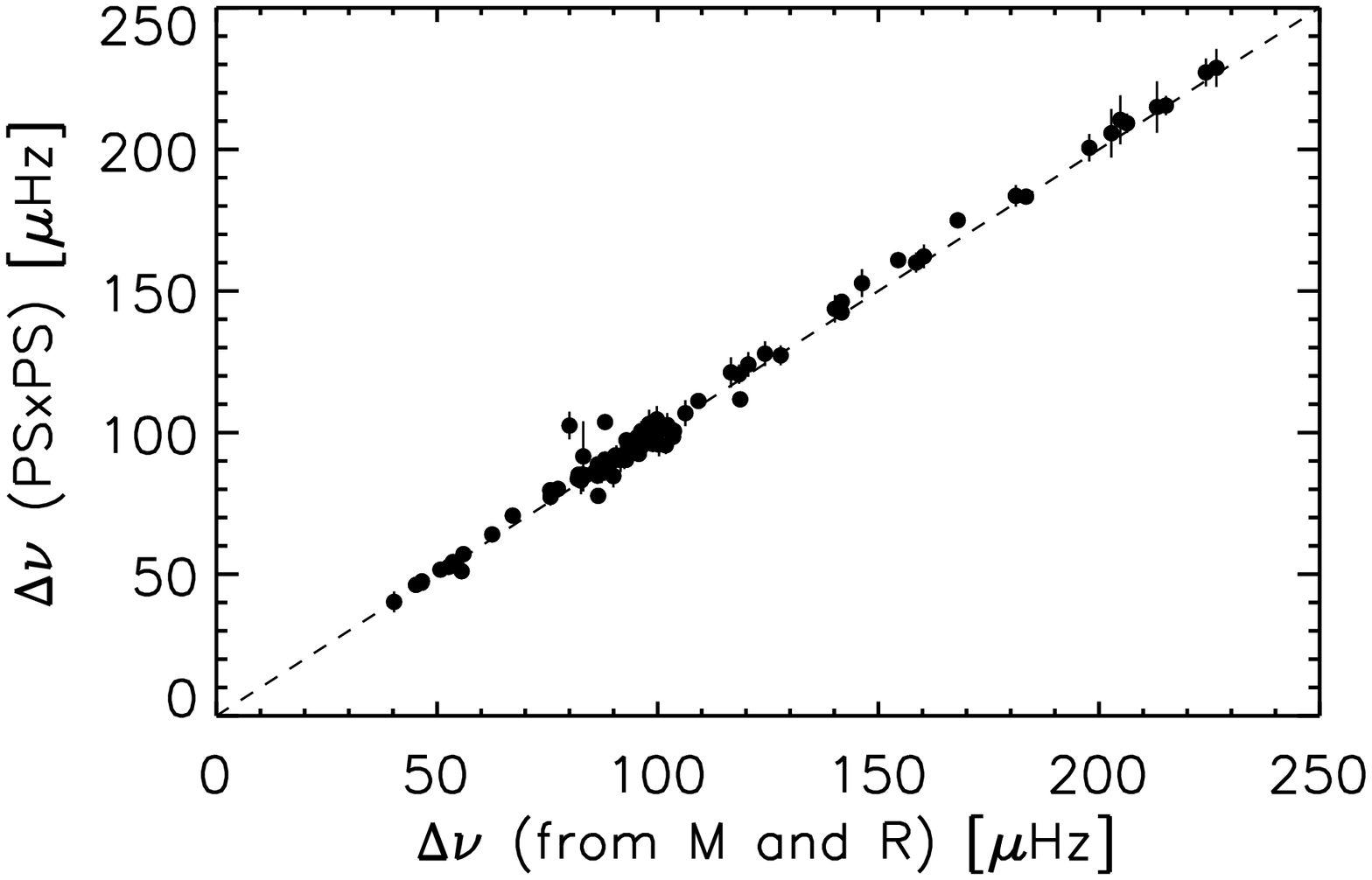}
\end{minipage}
\begin{minipage}{0.45\linewidth}
\centering
\includegraphics[width=\linewidth]{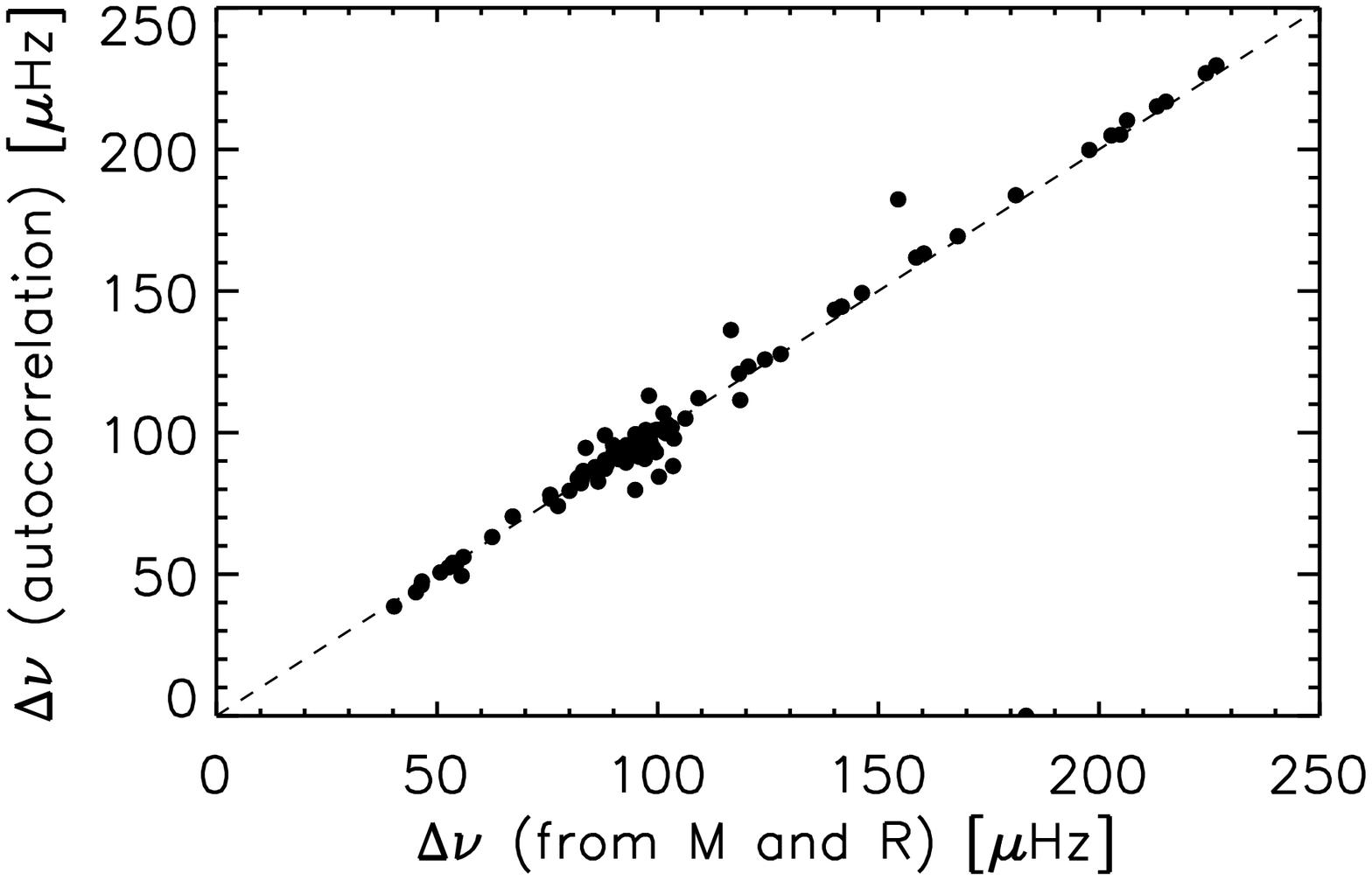}
\end{minipage}
\begin{minipage}{0.45\linewidth}
\centering
\includegraphics[width=\linewidth]{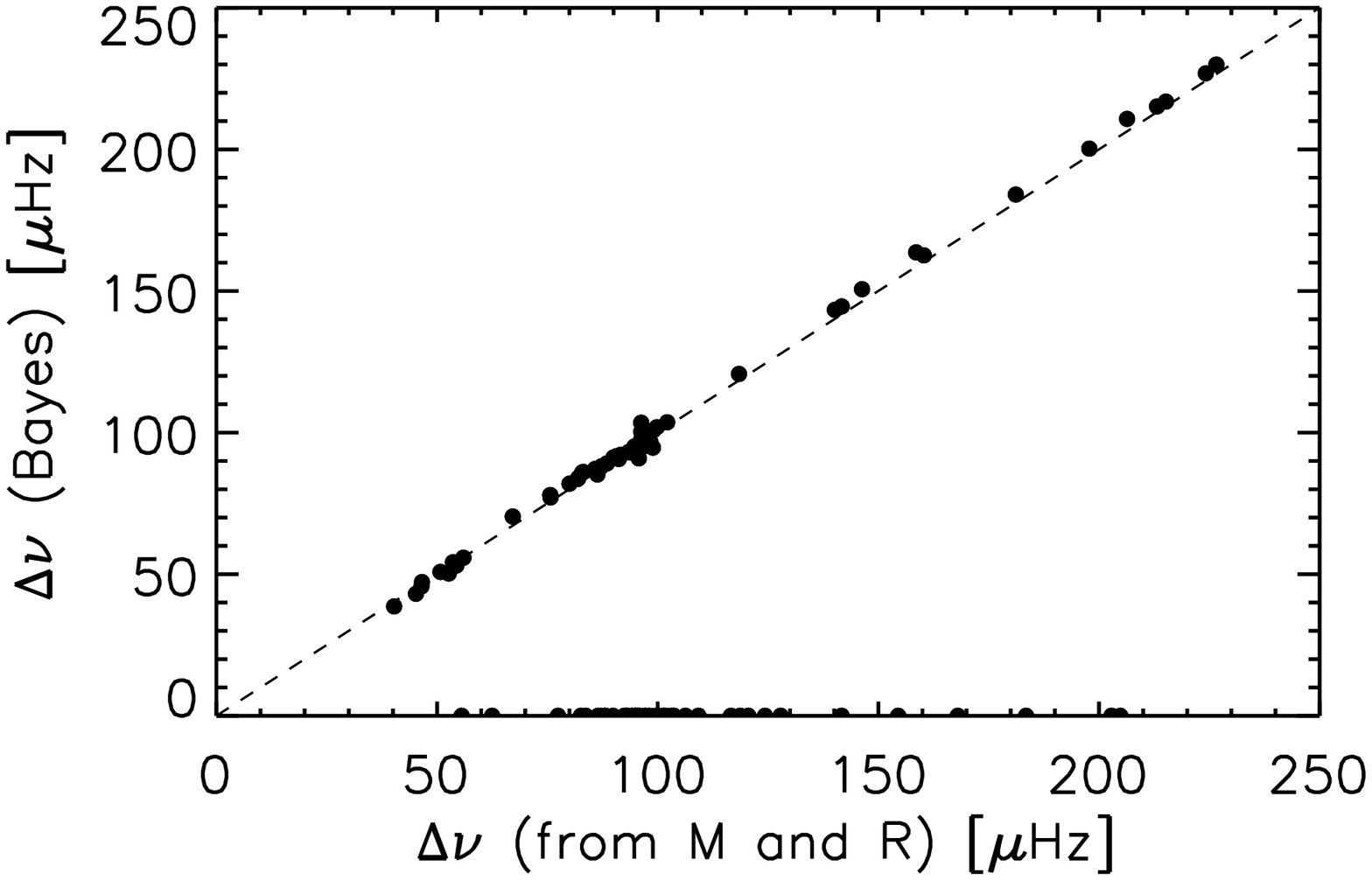}
\end{minipage}
\caption{{\sl Top:} $\Delta \nu$ results from the PS$\otimes$PS vs $\Delta \nu$ computed using the input $M$ and $R$ in Eq.~\ref{scale2}. In the left panel the $\Delta \nu$ values are computed as the power weighted mean centroids of the features in the PS$\otimes$PS. In the right panel $\Delta \nu$ values are computed using the Bayesian posteriory probability to compute the centroids of the features in the PS$\otimes$PS. {\sl Bottom:} $\Delta \nu$ results from autocorrelations vs $\Delta \nu$ computed using the input $M$ and $R$ in Eq.~\ref{scale2}. In the left panel the autocorrelations of the full oscillation frequency are used to compute $\Delta \nu$. In the right panel the individual frequencies obtained using a Bayesian approach are used to compute the autocorrelations, from which we determine $\Delta \nu$. The one-to-one relations are indicated with a dashed line.}
\label{dnumax}
\end{figure*}


For the input parameters we chose for $p_{\rm act}$ and $p_{\rm gran}$ the maximum power of the binned power spectrum and 0.001 times this value, respectively. Furthermore, the inputs for $\tau_{\rm act}$ and $\tau_{\rm gran}$ were 100\,000 and 1000 seconds, while the input value for $b$ was the mean power at high frequencies outside the oscillation range. As a first estimate, we chose $a$ to be equal to 2. To obtain the optimal fit, we vary the input parameters slightly. We randomly select one of the fitting parameters and multiply this by:
\begin{equation}
1 + 0.3 \times rn,
\end{equation}
where, $rn$ denotes a random number that has a normal distribution with a mean of zero and standard deviation of one. To minimise boundary effects from the excluded oscillation frequency range we vary this range by extending or shortening it on both sides by 1/6 of the length of the original frequency range. After repeating this 200 times for each excluded oscillation range, the fit with the lowest $\chi^2$ is used as the best-fitting background fit.

In a few cases, the fitting with two power laws does not work properly. This is because only one decaying profile is visible in the data, either due to the presence of the oscillations at the same frequency as the hump of the second decaying profile, or due to a too low a signal-to-noise ratio, i.e., high white noise or low signal, or a combination of the two. In these instances we fit only for one power law and the offset $b$. This does not provide us with an optimal fit at low frequencies (below $\sim$ 10 $\mu$Hz), and the parameters cannot be used to infer properties of granulation and activity. However, at higher frequencies ($>$ 100 $\mu$Hz), where the oscillations reside the single power-law fit provides a reasonable estimate of the background. The standard deviations of the fitting parameters are used as errors. 

\subsection{Average large separation} For the estimation of the large separation ($\Delta \nu$) we compute the PS$\otimes$PS in the frequency range of the oscillations. Here, we take into account that $\Delta \nu$ depends on frequency and compute the PS$\otimes$PS of a power spectrum with a stretched frequency axis. Determining $\Delta \nu$ from the stretched power spectrum provides a more reliable measure of $\Delta \nu$ and if required an estimate of the gradient of the large spacing with $n$, $\delta \Delta \nu$/$\delta$$n$ can be made. For more details on the stretching see the last paragraph of Section 3.1. The derivation of the gradient of the large separation is presented in Appendix A. In the PS$\otimes$PS (see Fig~\ref{PSPS}) we determine the position of the $\Delta \nu$/2 and $\Delta \nu$/4 features. The centroids and uncertainties of these features are computed in two ways. In the first method, we determine the power weighted centroids of the features in the PS$\otimes$PS and their errors are computed as the standard deviation of grouped data (see Eq.~\ref{errdnu} in Appendix A).
In a second method we compute the Bayesian posterior probability of the points in the PS$\otimes$PS, using the same equations as for the individual frequencies, i.e., Eqs.~\ref{post} - \ref{H1}, which are discussed in Section 3.6 (see also \citet{broomhall2009} and \citet{appourchaux2009}. Using these probabilities we compute the posterior weighted centroid of the feature. The interval with a probability of the feature not being due to noise higher than 68.27\%, i.e., 1$\sigma$ in a Gaussian distribution, is used as the uncertainty interval.
Finally, for both methods, we determine $\Delta \nu$ by computing a weighted average of $\Delta \nu$/2 and $\Delta \nu$/4. 

We also compute $\Delta \nu$ from an autocorrelation of the full oscillation frequency range and from an autocorrelation using the individual oscillation frequencies determined with the Bayesian approach only (again, see Section 3.6). A Gaussian is fitted to the feature at $\Delta \nu$ in the respective autocorrelations and the width of the Gaussian is used as an estimate of the error.

\subsection{Maximum mode amplitude, amplitude envelope and frequency of maximum amplitude} 
Our next package provides estimates of the maximum mode amplitude, and the mode-amplitude envelope as a function of frequency. Our results are scaled to be equivalent radial-mode amplitudes.

In summary, for method I we began by subtracting the background fit from the power spectrum. The resulting, residual power spectrum is averaged over the range occupied by the modes using a boxcar filter of width $3\Delta \nu$. Next, we multiply this averaged, residual spectrum by the large frequency spacing $\Delta\nu$, and finally we divide by a constant factor, $c$, to allow for the effective number of modes in each slice $\Delta\nu$ of the spectrum. The value of $c$ is chosen so that the above procedure gives observational estimates, as a function of frequency, of the power envelope for radial modes. $c$ is computed assuming the presence of 4 frequencies in each $\Delta \nu$ interval with $\ell$ = 0, 1, 2, 3, with relative power per mode of 1.0, 1.5, 0.5, 0.03 respectively. $c$ is the total power we expect in a $\Delta \nu$ interval, i.e., 3.03. There is a slight dependence of $c$ on limb darkening and thus on $T_{\rm eff}$ and $\log$ $g$, but the values change only by a few percent and we ignore those changes here.

The highest value of the power envelope is an estimate of the maximum mode power. The mode amplitude envelope and the maximum mode amplitude are given by the square root of the power envelope, and square root of the maximum power, respectively. The frequency at which the maximum mode power occurs is $\nu_{max}$.

We choose to average the spectrum using a boxcar filter as opposed to the Gaussian filter (of width $4\Delta\nu$) adopted by \citet{kjeldsen2008}. This is because it allows us to estimate very
straightforwardly uncertainties for the amplitudes, here in independent frequency ranges of width $3\Delta\nu$. For example, we estimate the uncertainty of the maximum mode power as the standard
deviation of the powers in each bin in the frequency range 3$\Delta \nu$ that contribute to the estimated maximum power. We may then calculate independent averages in ranges on either side of the maximum, to give an estimated power envelope with uncertainties. Uncertainties for the amplitudes follow by remembering that fractional errors on the mode amplitudes are equal to half those
on the mode powers.

Our decision to average over 3$\Delta \nu$ was to some extent determined by the following obvious compromise. The narrower the range, the more we avoid smoothing out potentially interesting features of the amplitude envelope, while the wider the range, the less subject we are to fluctuations due to the stochastic nature of the modes. But there is also another important factor, which argues for adopting a wider range: that is to suppress biases in the estimated maximum mode amplitudes when the signal-to-noise ratio is quite low.


The frequency at which maximum oscillation power occurs ($\nu_{\rm max}$) is computed as the weighted mean frequency of the oscillation power with the error computed from the standard deviation of grouped data (see Eq.~\ref{errdnu}).


In a second method (hereafter method II), we fit a Gaussian to the binned oscillation power, where the binning is performed over intervals of 2$\Delta \nu$. The height of the Gaussian fit is then converted to amplitude per radial mode by multiplying by $\Delta \nu$/$c$, as per the other approach. We use the standard deviation of the fit parameters to compute the errors. 
The centre of the Gaussian fit is $\nu_{\rm max}$. 

\subsection{Linewidth of most prominent modes}
We seek a straightforward and robust method for determining from the power spectrum the linewidth shown by the most prominent modes.

Our method relies on the fact that the height in the power spectrum of a solar-like (i.e., damped) mode peak depends not only on the total power of the mode, but also (crucially to our method here) on the
linewidth (or equivalently the damping time) of the mode and the intrinsic resolution in frequency of the spectrum. The height, $H$, in units of power per Hertz is well described in both the resolved and unresolved regime by \citep{fletcher2006,chaplin2009}:
 \begin{equation}
 H(T) = \frac{2A^2T}{\pi T \Delta + 2},
 \label{linewidth}
 \end{equation}
where $A^2$ is the total power of the mode, $\Delta$ is the FWHM linewidth of the mode peak, and $T$ is the effective length of the observations. We may re-express Equation~\ref{linewidth} in terms of the intrinsic (or
natural) resolution in frequency $\delta = 1/T$. Substitution and subsequent re-arrangment of the equation then gives the following
 \begin{equation}
 \frac{2A^2}{\pi H(\delta)} = \Delta + \frac{2\delta}{\pi},
 \label{deltaprime}
 \end{equation}
which is the form required to explain our method. For a range of values of $\delta$ we estimate the ratio $A^2/H(\delta)$ of the most prominent radial mode in the spectrum (as explained in the next paragraph
below). A plot of $A^2/H(\delta)$ versus $\delta$ then yields data following a linear relationship. We fit a straight line to the data, and the intercept on the ordinate in principle provides an estimate of the linewidth, $\Delta$. Evaluation of the spectrum at different $\delta$ is achieved by averaging the spectrum of the full timeseries over different numbers of bins $M$ (thereby degrading the intrinsic
resolution as required). If $T$ is taken to be the effective length of the timeseries, this means that $\delta = M/T$.

We estimate the ratio $A^2/H(\delta)$ as follows. We already have an estimate of $A^2$ courtesy of the mode amplitude package in Section 3.4. To estimate the heights $H(\delta)$ in each $M$-bin-averaged spectrum we simply take the highest power spectral density in the range $\Delta\nu/2$ about $\nu_{\rm max}$. These estimates are only a proxy of the true, underlying $H(\delta)$, which means that to correctly estimate
linewidths $\Delta$ we must apply an empirical correction to the results. We found from simulations that a linear correction with both an offset and slope of 0.4 applied to the raw estimates of the linewidth is sufficient for this purpose. 

Our simple proxy of $H(\delta)$ may sometimes have been estimated from the most prominent $\ell=1$ mode (depending on the inclination of the star), when it is the height of the most prominent $\ell=0$ mode that we require. In our first version of the pipeline, we accept this potential uncertainty, and note that its main effect will be to add some additional scatter to the results. Any bias is taken care of by
the empirical correction above.


\begin{figure}
\begin{minipage}{\linewidth}
\centering
\includegraphics[width=\linewidth]{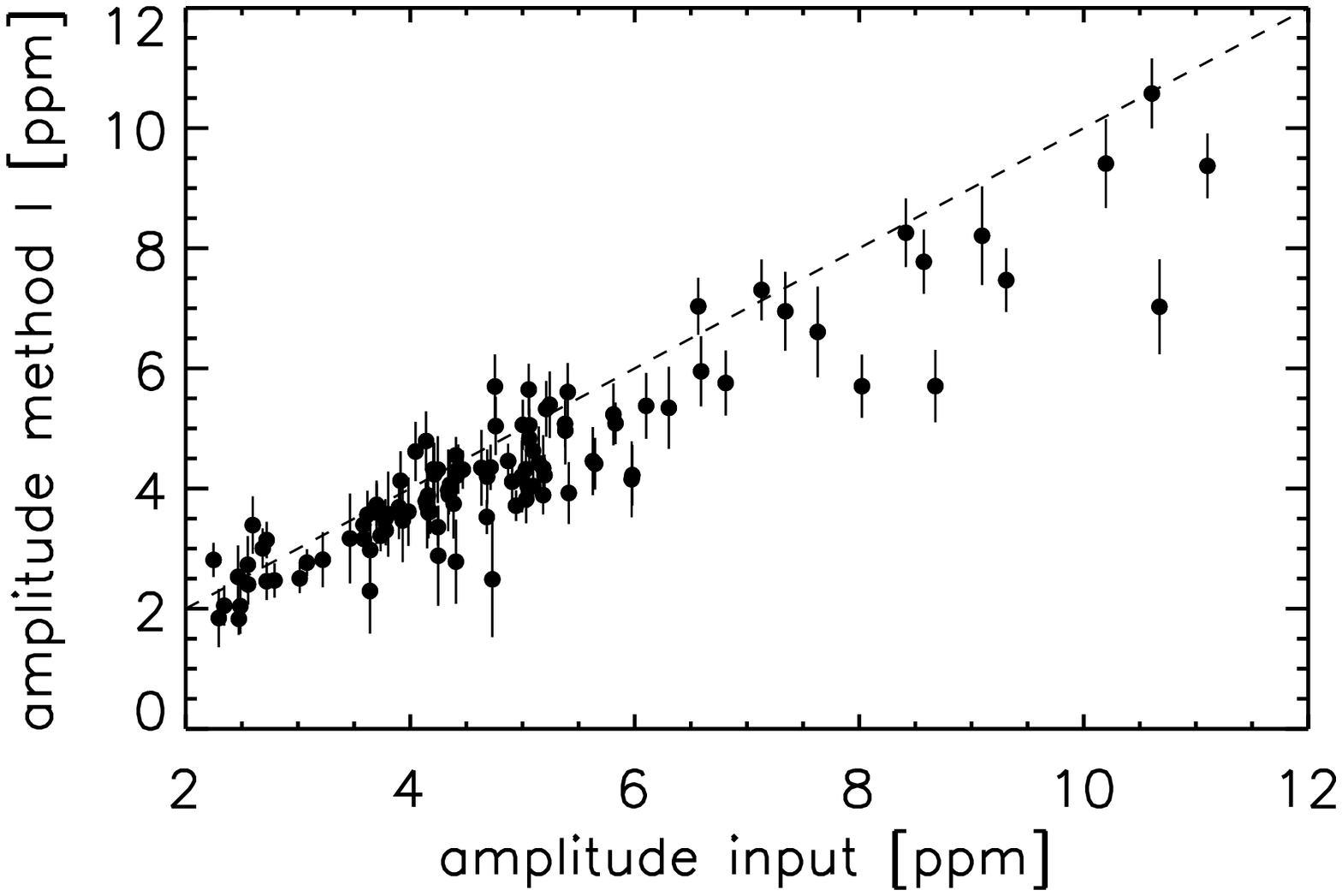}
\end{minipage}
\begin{minipage}{\linewidth}
\centering
\includegraphics[width=\linewidth]{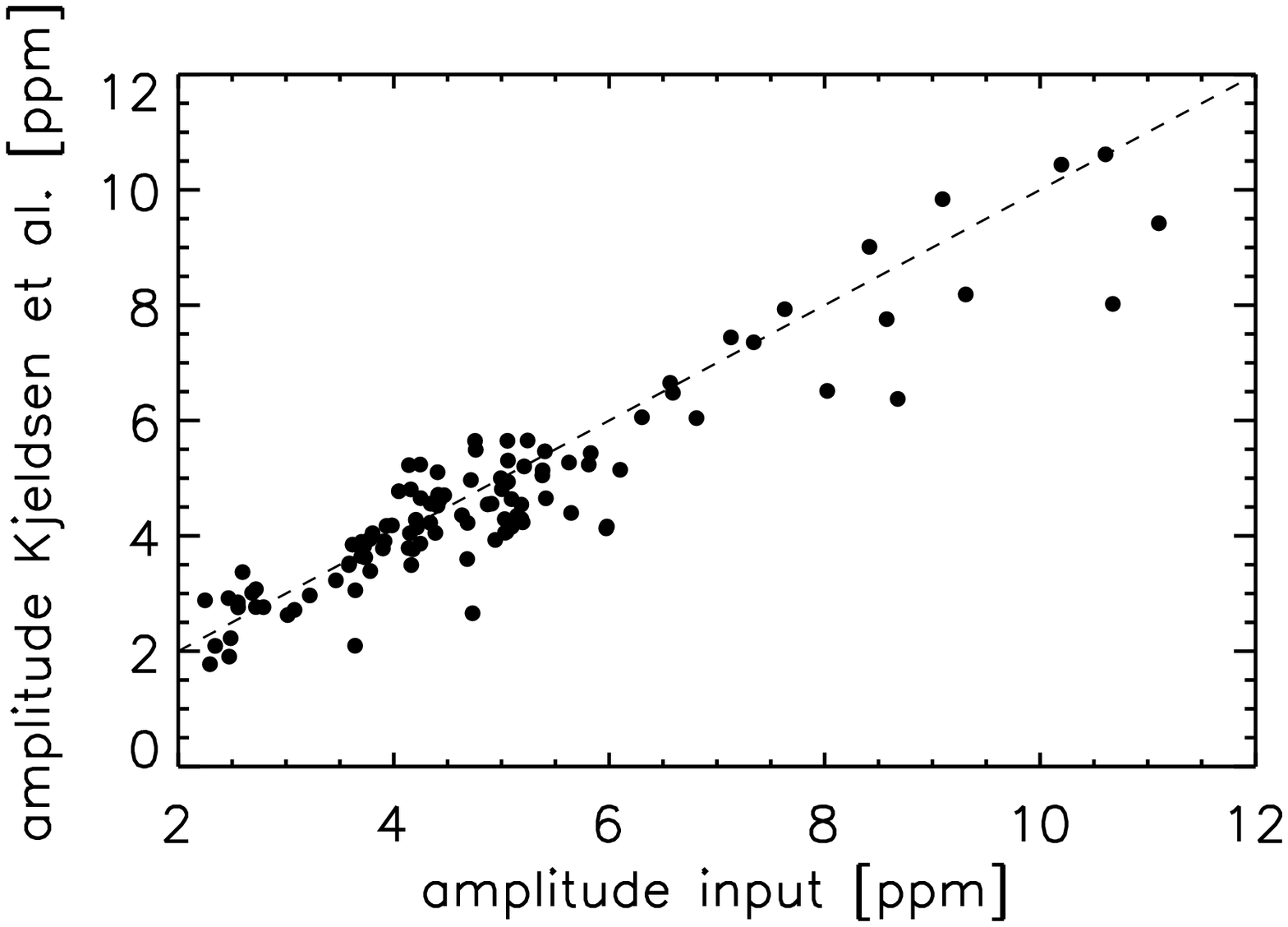}
\end{minipage}
\begin{minipage}{\linewidth}
\centering
\includegraphics[width=\linewidth]{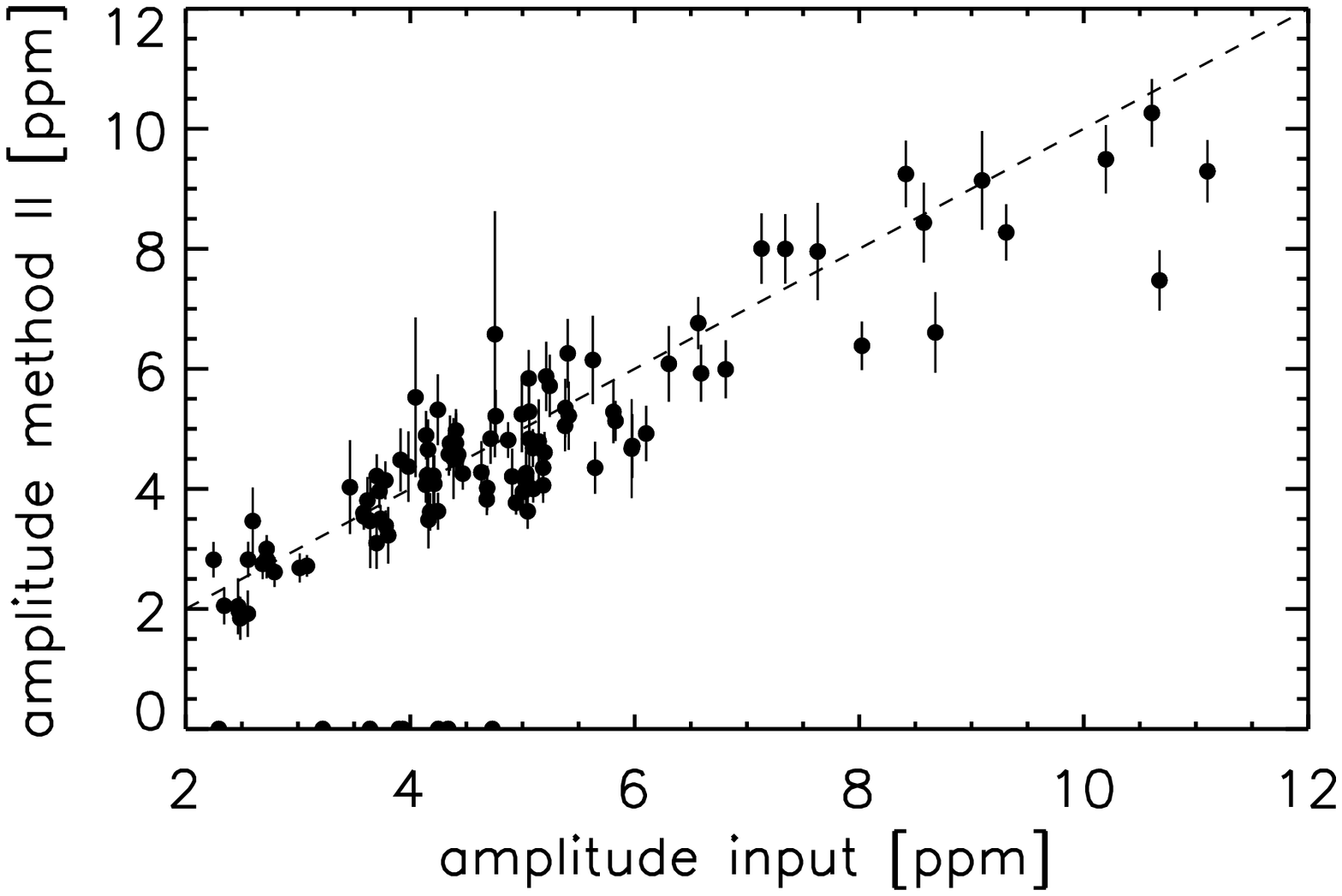}
\end{minipage}
\caption{Results for the average mode amplitude from method I (top), \citet{kjeldsen2008} (centre) and method II (bottom) vs. the input mode amplitude. The one-to-one relations are indicated with a dashed line.}
\label{compareampl}
\end{figure}

\subsection{Individual frequencies}
For the determination of  individual frequencies we used a Bayesian approach adapted from \citet{broomhall2009} and references therein. We want to test whether the power at each frequency in the power spectrum could be the result of a component of a stochastically excited mode ($H_1$ hypothesis) or due to noise ($H_0$ hypothesis).

As explained by \citet{appourchaux2009} we aim to compute the posterior probability of $H_0$ ($p(H_0|x)$) given the observed data $x$, i.e.,
\begin{equation}
p(H_0|x)=\left(1+\frac{p(x|H_1)}{p(x|H_0)}\right)^{-1},
\label{post}
\end{equation}
where we assumed that both hypotheses are, a priori,  equally probable.
For the $\chi^2$ 2 dof statistics of the power spectrum, the probability of observing $x$ given that there is only noise, i.e., the probability of observing $x$ if the $H_0$ hypothesis is true, ($p(x|H_0)$) is:
\begin{equation}
p(x|H_0)=e^{-x},
\label{H0}
\end{equation}
where $x$ is the observed power divided by the background. 

For the alternative hypothesis $H_1$, i.e., the probability of observing $x$ given that there is signal, we assume that we do not know a priori the mode height $H$. Therefore, we assume that the height can be taken from a uniform distribution between 0 and $H_s$ (see Appendix A for details on the determination of $H_s$). We can then compute the probability of observing $x$ if the $H_1$ hypothesis is true ($p(x|H_1)$) as follows \citep{appourchaux2009}:
\begin{equation}
p(x|H_1)=\frac{1}{H_s} \int^{H_s}_0 \frac{1}{1+h} e^{-x/(1+h)}dh.
\label{H1}
\end{equation}
Frequencies with a posterior probability less than 0.5\% are taken to be candidate oscillation frequencies. The 0.5\% posterior probability was chosen based on 1000 Monte Carlo simulations performed with the asteroFLAG code \citep{chaplin2008} of solar analogs, for which we computed the fraction of real detections for different posterior probabilities (see Fig.~\ref{postprob}). The final frequencies ($\nu_{\rm final}$) were computed using the parameter estimation:
\begin{equation}
\nu_{\rm final} = \frac{ \int \nu p(H_0|x) d \nu}{\int p(H_0|x) d \nu},
\label{avgfreq}
\end{equation}
For the integration range, we use the frequency range for which we found that the posterior probability to find signal was larger than 68.27\%, i.e, 1$\sigma$ in a Gaussian distribution. We also used this interval as the estimated error. We tested this error estimation by performing 1000 Monte Carlo simulations of one single stochastically excited mode with flat background noise. For each simulation we computed the final frequency and its error. Then we expressed the offset between the computed frequency and input frequency in terms of its error. We see that for 77\% of the tests the offset is within 3$\sigma$ and for 89\% of the tests it is within 5$\sigma$.

In cases where more than three significant oscillation frequencies could be detected,  we used these individual frequencies to compute the large separation from the autocorrelation of the frequencies.

\subsubsection{Small separations} We have investigated for how many stars we might be able to find the small separation ($\delta \nu_{02}$), i.e., for how many stars we could see more than two ridges in the \'{e}chelle diagram. This was the case for less than 10\% of the stars. This low number is most likely caused in part by the fact that we set the threshold posterior probability at only 0.5\%, which reduces the false alarm rate, but also the number of identified frequencies. Therefore, for most main-sequence stars only two ridges are present in the \'{e}chelle diagram. Because of the low percentage of stars for which we might be able to identify the small separation with the strict thresholds currently applied, we do not include such computation in the automated pipeline presented here.  In a further analysis using peak-bagging techniques, $\delta \nu_{02}$ will be obtained. These results will be presented by Fletcher et al. (2009), in preparation.
\newline


\begin{figure}
\begin{minipage}{\linewidth}
\centering
\includegraphics[width=\linewidth]{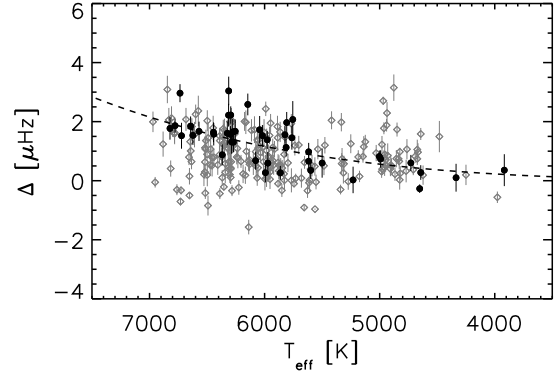}
\end{minipage}
\caption{Line width as a function of temperature, with results for stars brighter than 9$^{\rm th}$ mag indicated with black dots. Results for fainter stars are indicated with grey diamonds. The $\Delta$ $\sim$ $T_{\rm eff}^4$ input relation \citep{chaplin2009a} is indicated with a dashed line.}
\label{comparewidth}
\end{figure}

\section{Results}
We are able to detect oscillations in 260 out of the 353 artificial stars, i.e, nearly 75\%. For these stars we estimated the oscillation parameters and background as described in the previous section, which we then compared with the input values used to create the artificial data. 

\begin{figure*}
\begin{minipage}{0.45\linewidth}
\centering
\includegraphics[width=\linewidth]{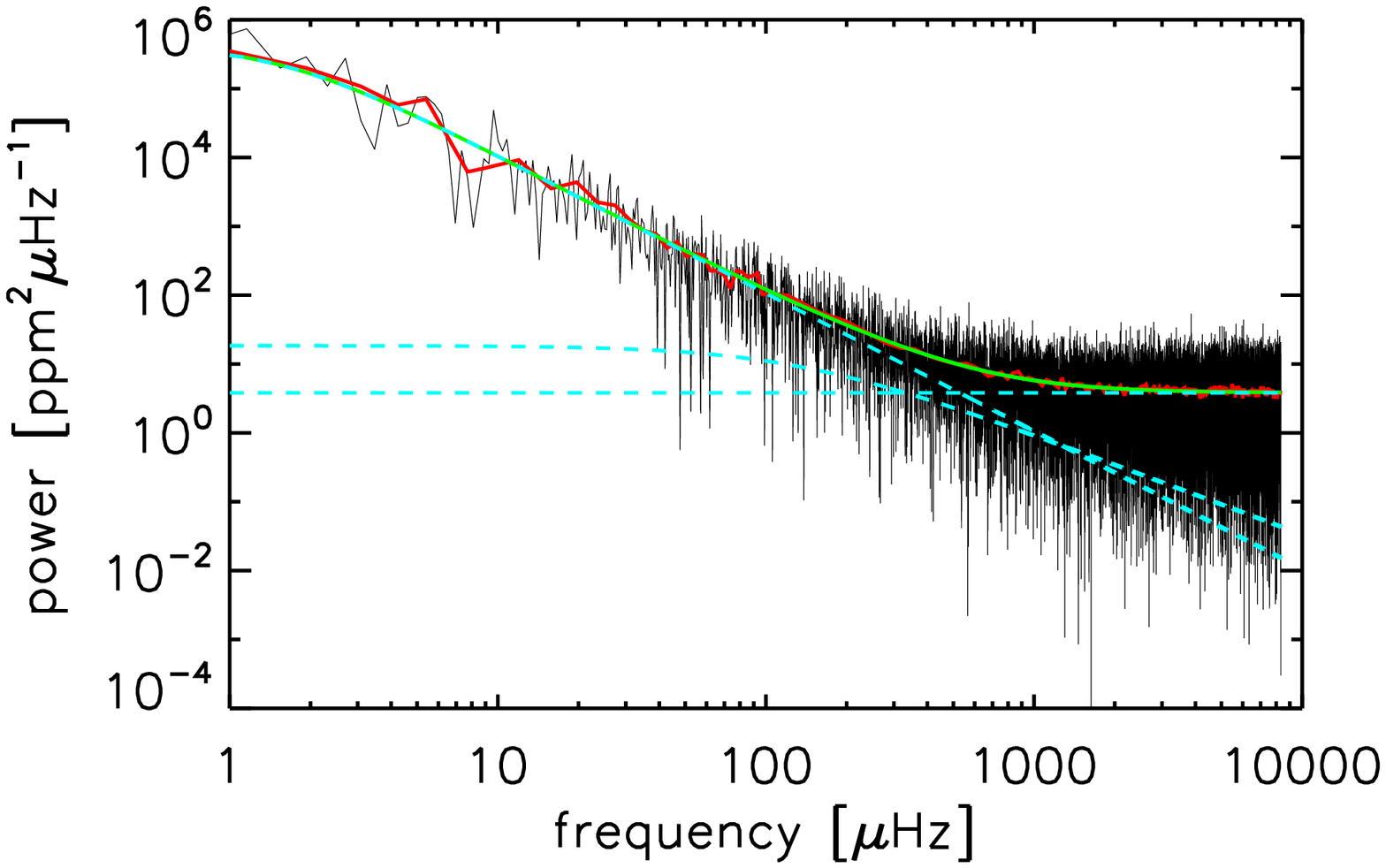}
\end{minipage}
\begin{minipage}{0.45\linewidth}
\centering
\includegraphics[width=\linewidth]{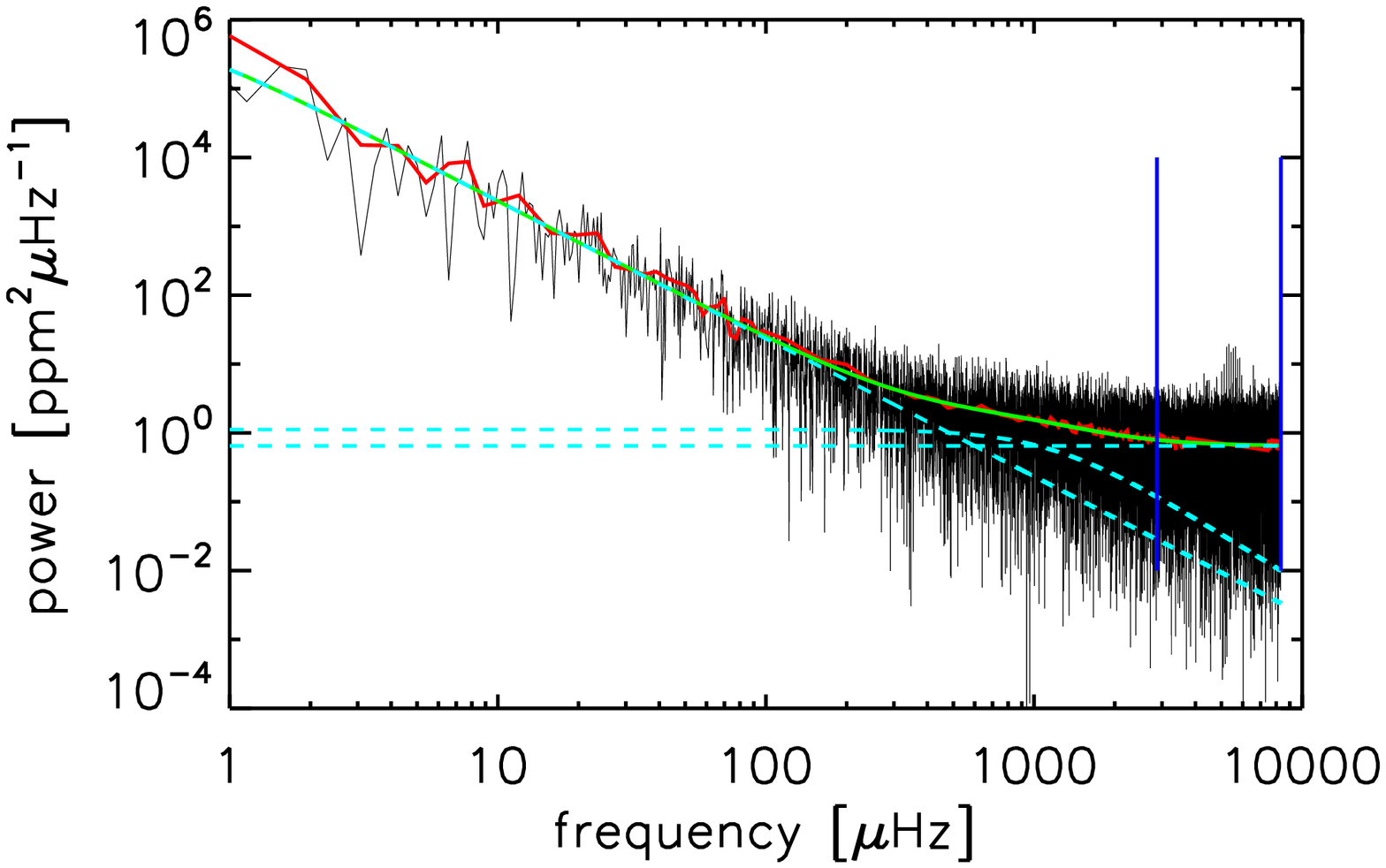}
\end{minipage}
\begin{minipage}{0.45\linewidth}
\centering
\includegraphics[width=\linewidth]{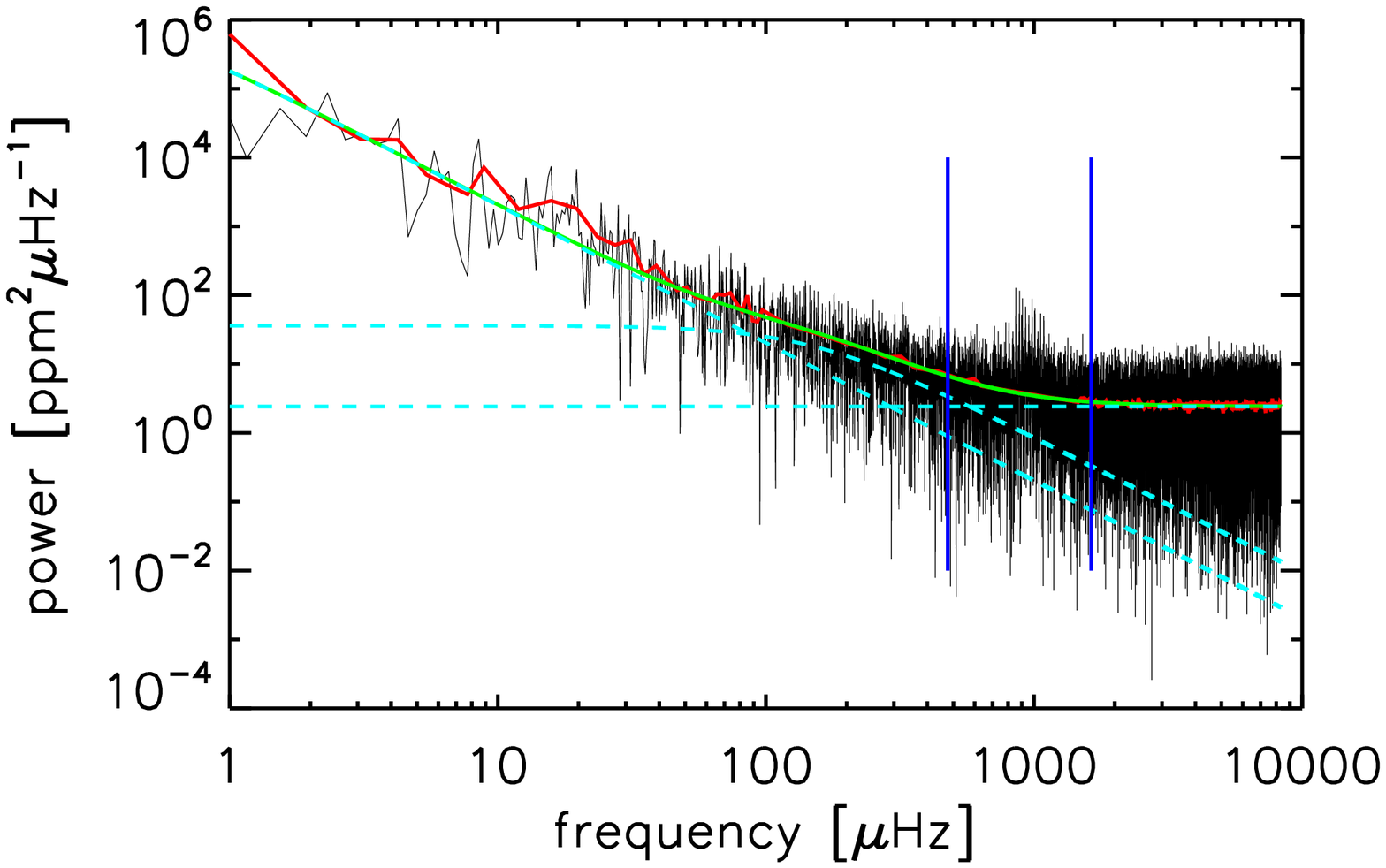}
\end{minipage}
\begin{minipage}{0.45\linewidth}
\centering
\includegraphics[width=\linewidth]{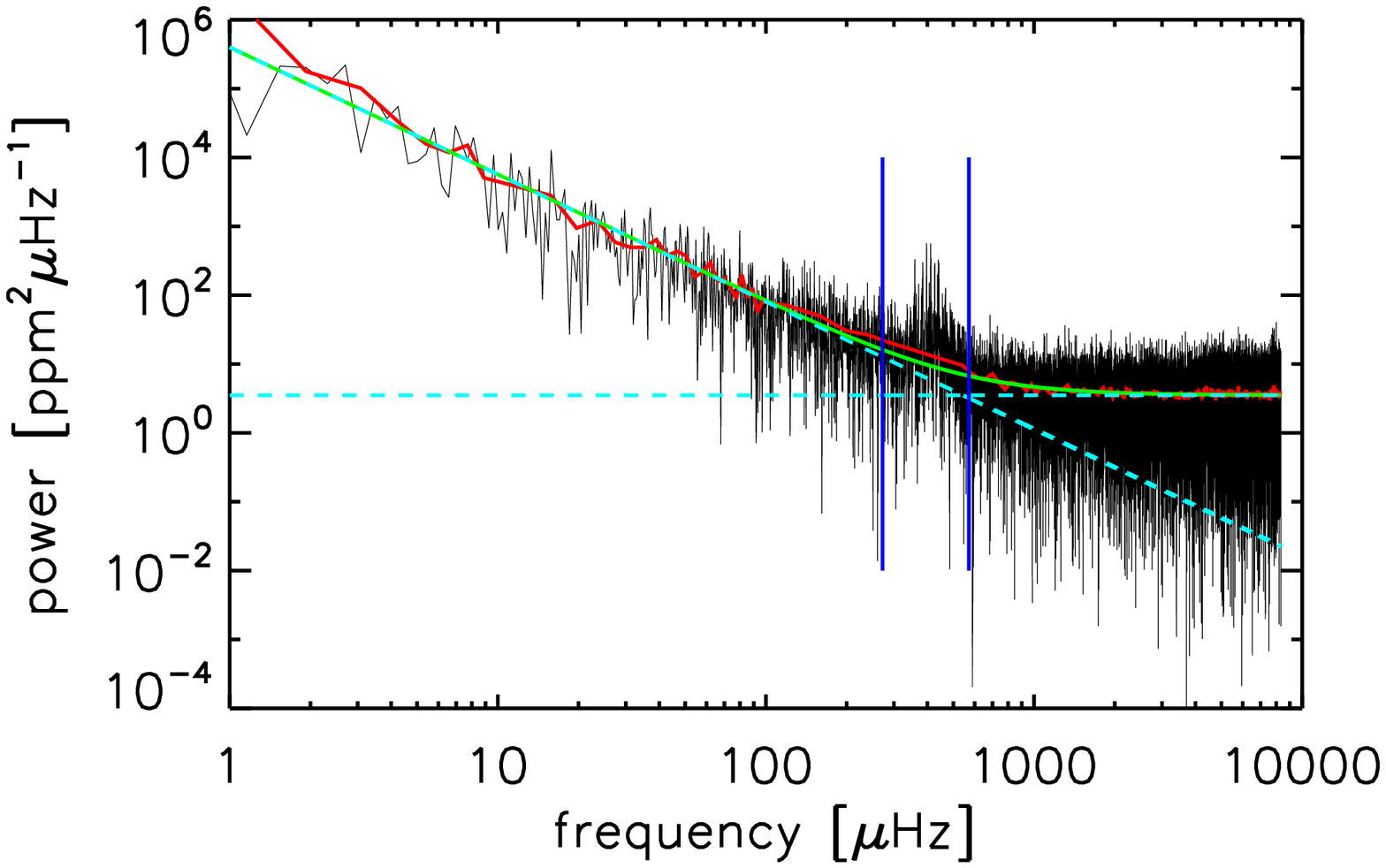}
\end{minipage}
\caption{Power spectra of 4 artificial Kepler stars with oscillations at different frequencies, displayed in a log-log scale. The red lines indicate the binned power spectrum and the green lines the background fits, with the individual components shown in cyan dashed lines. The identified oscillation frequency ranges are indicated with vertical blue lines. No oscillations could be identified in the top left power spectrum, while for the bottom right power spectrum we could only fit one power law for the background.}
\label{bgfits}
\end{figure*}

\begin{figure}
\begin{minipage}{\linewidth}
\centering
\includegraphics[width=\linewidth]{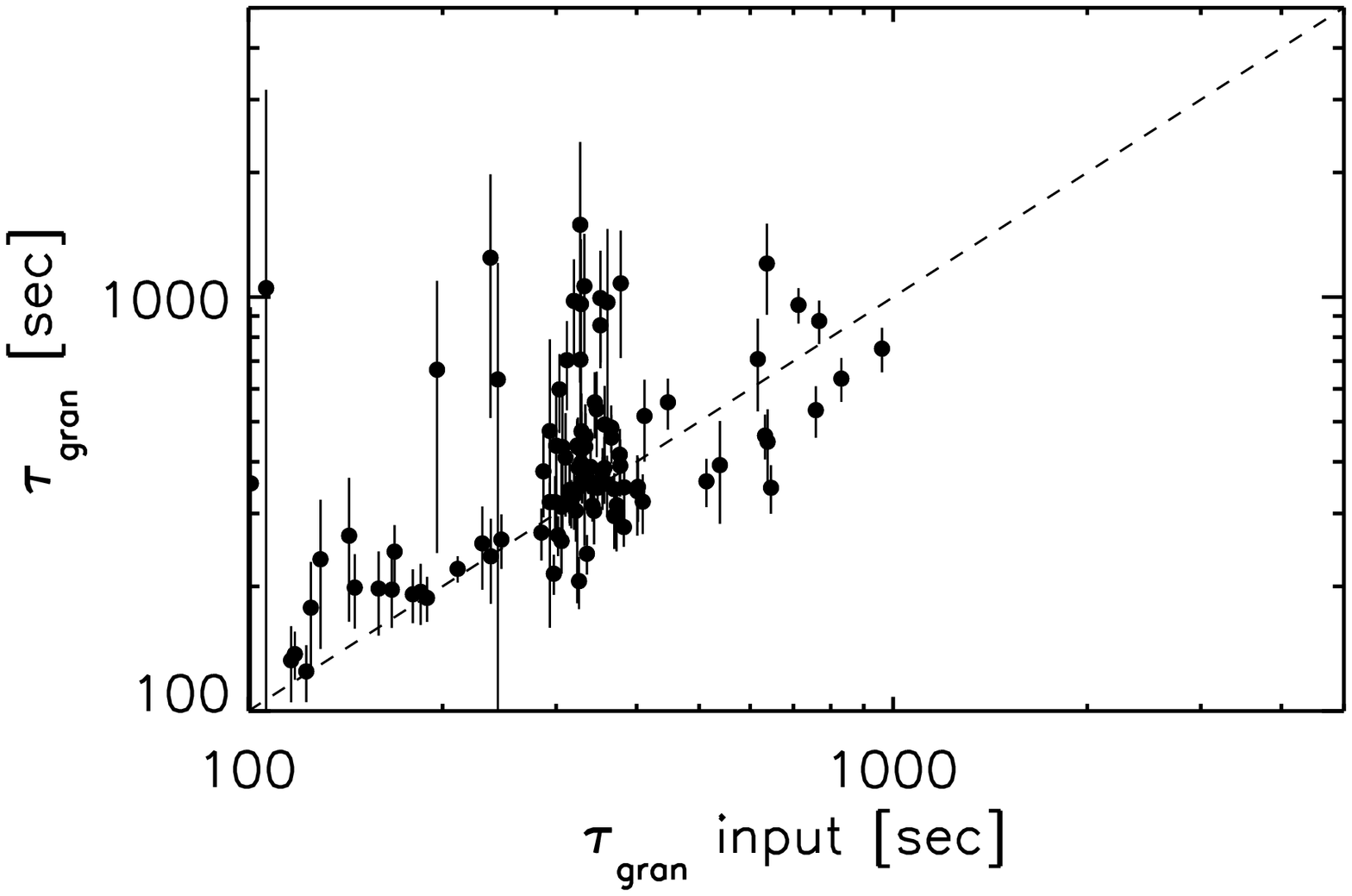}
\end{minipage}
\begin{minipage}{\linewidth}
\centering
\includegraphics[width=\linewidth]{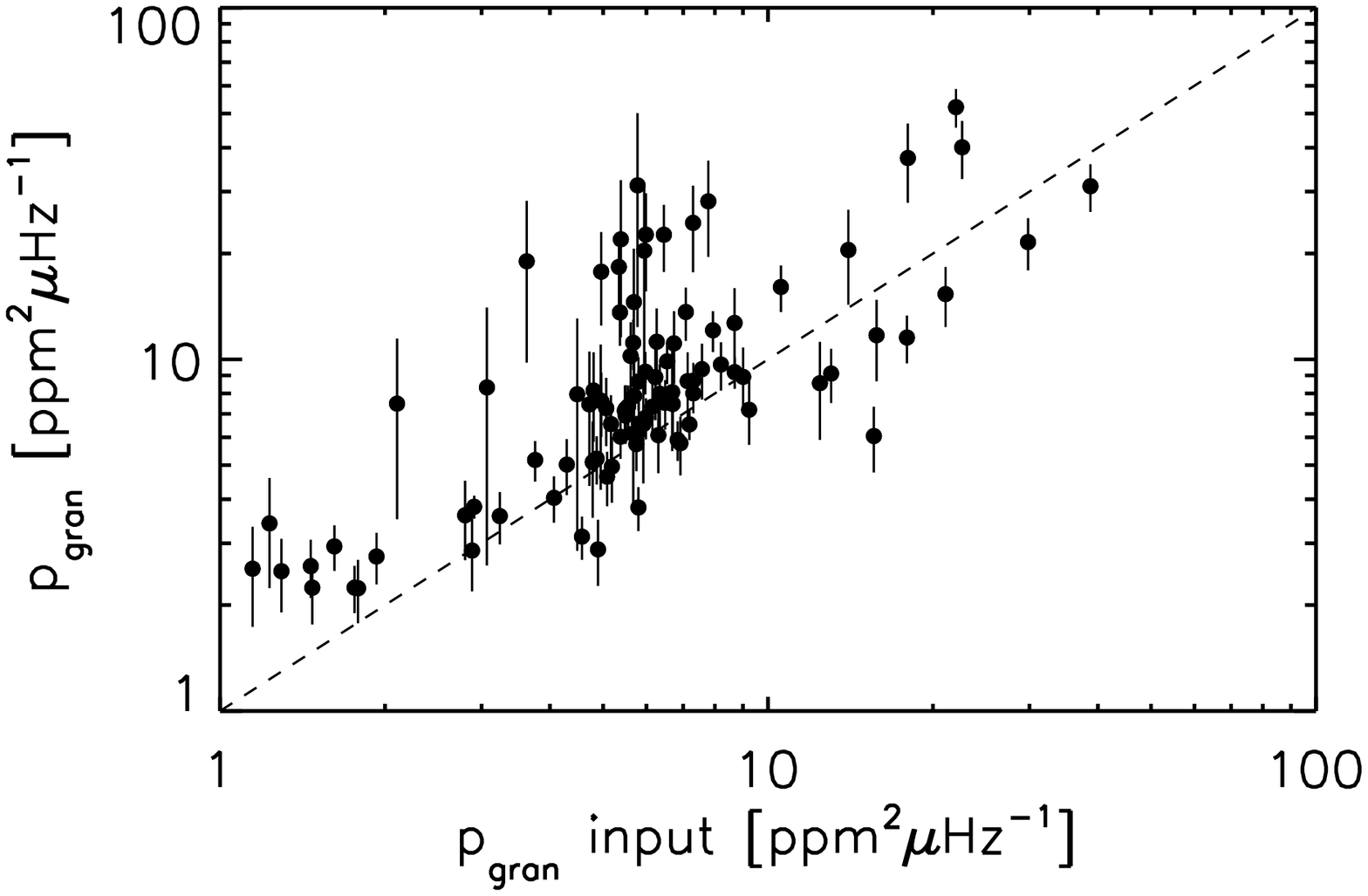}
\end{minipage}
\caption{Results of the timescale ($\tau_{\rm gran}$, top) and power ($p_{\rm gran}$, bottom) of the granulation as a function of the input values. The one-to-one relations are indicated with a dashed line.}
\label{gran}
\end{figure}

\subsection{Oscillation parameters}
In Figs.~\ref{numax} and \ref{dnumax} we compare our results for $\nu_{\rm max}$ and $\Delta \nu$ with the values computed from the input mass ($M$), radius ($R$) and effective temperature ($T_{\rm eff}$), using the scaling relations \citep{kjeldsen1995}:
\begin{equation}
\nu_{\rm max} = \frac{M/M_{\odot}}{(R/R_{\odot})^2\sqrt{T_{\rm eff}/5777K}}3050\mu \rm Hz,
\label{scale1}
\end{equation}
\begin{equation}
\Delta \nu = \sqrt{\frac{M/M_{\odot}}{(R/R_{\odot})^3}}134.9 \mu \rm Hz.
\label{scale2}
\end{equation}

\begin{figure}
\begin{minipage}{\linewidth}
\centering
\includegraphics[width=\linewidth]{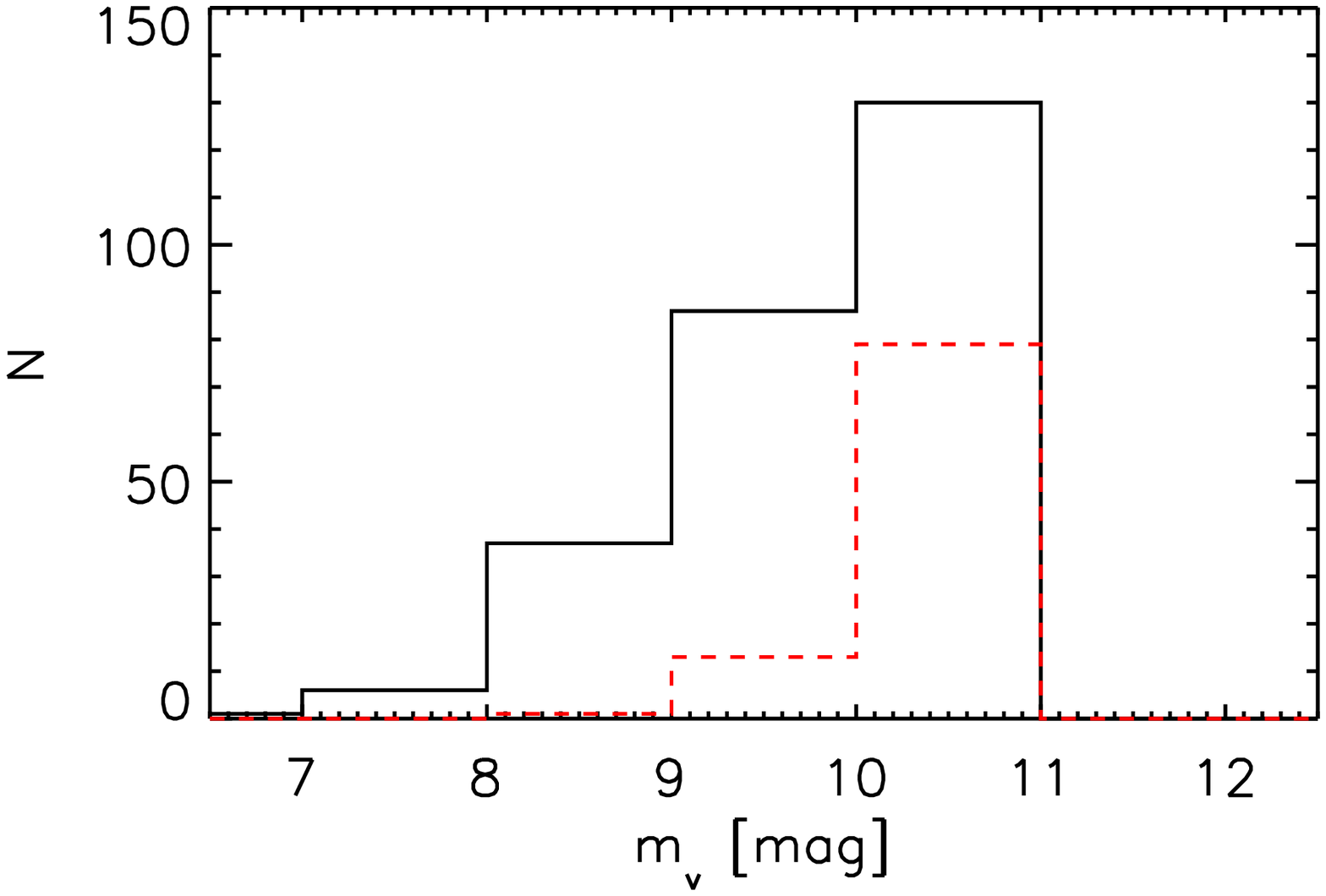}
\end{minipage}
\caption{Distribution of apparent magnitudes of stars for which we detected oscillations (black solid line) and for stars for which we did not detect oscillations (red dashed line).}
\label{mag}
\end{figure}

The values obtained for both $\nu_{\rm max}$ (Fig.~\ref{numax}) and $\Delta \nu$ (Fig.~\ref{dnumax}) are in good agreement with the input values, for each of the implemented methods, although a slight overestimation is present in the determined $\nu_{\rm max}$ values from method I at higher frequencies where the height of the oscillations decreases \citep{chaplin2009a}. For method I, 54\% of our $\nu_{\rm max}$ values agree within uncertainties with the input values, while this percentage increases to 97\% within 3 times the computed uncertainties. For method II, we find that 37\% of our $\nu_{\rm max}$ values agree with the input values within their uncertainties, which increases to 91\% agreement within 3 times the uncertainties. In 96\% of the stars for which we could detect oscillations we found $\nu_{\rm max}$ with both methods.

For $\Delta \nu$ we found that 93\%, 90\%, 86\% and 97\% of our values agree with the input values within 5\%, for results computed with the weighted mean of the features in the PS$\otimes$PS, the Bayesian probabilities in the PS$\otimes$PS, autocorrelation of full oscillation frequency range and autocorrelation of individual frequencies, respectively. The uncertainties for $\Delta \nu$ computed with the Bayesian probabilities in the PS$\otimes$PS are larger and seem more realistic than for all other methods. For this method 94\% of our values agree with the input within 3 times the uncertainties, while this is only 44\%, 20\% and 7\% for the other methods, i.e., weighted mean of the features in the PS$\otimes$PS, autocorrelation of full oscillation frequency range and autocorrelation of individual frequencies, respectively. For the first three methods we find $\Delta \nu$ in all stars for which we detect oscillations. For $\Delta \nu$ computed with the individual frequencies we have results for 60 \% of the stars. Due to the better error estimate the results of the Bayesian probabilities in the PS$\otimes$PS are most reliable. Despite their underestimated errors, the $\Delta \nu$ values computed with the other methods are in more than 90 \% of the cases compatible (within 3 times the uncertainties) with the Bayesian values.

In Fig.~\ref{compareampl} our results for the maximum amplitude per radial oscillation mode are shown as a function of the input maximum amplitude per radial mode. For comparison we also computed estimates of the maximum amplitude using the method of \citet{kjeldsen2008}. The results are consistent with the one-to-one relation, and for each method $\sim$~90\% of our amplitude values are consistent with the input values within 3 times the computed uncertainties. 

Also, we computed the width of the frequency peaks in the power spectrum. The results are shown in Fig.~\ref{comparewidth}. The input values of the line width follow the relation $\Delta$~$\sim$~$T_{\rm eff}^4$, and we see in Fig.~\ref{comparewidth} that for stars brighter than 9$^{\rm th}$ magnitude (black dots) our results for $\Delta$ are qualititatively in agreement with this relation. Good signal-to-noise ratio is required for this method and for fainter stars the scatter in the results becomes considerably larger.

Four examples of background fits to power spectra with oscillations at different frequencies are shown in Fig.~\ref{bgfits}, all of which have a $\chi^2$ of the order of 1. We also compare in Fig.~\ref{gran} our fitted values of the granulation parameters $p_{\rm gran}$ and $\tau_{\rm gran}$ with the artificial input values. Here, we see that the results follow the one-to-one relation with the input value, but with non-negligible scatter. 


Note that for stars where a detection was made, we were not always able to determine the oscillation parameters with all methods described in Section 3. For each parameter we have at least one method that produces a result for all stars with detected oscillations, but for some methods we have results for fewer stars, down to 60\% of the stars with detected oscillations. The quoted percentages are always computed for stars with a result for the considered method. 

\subsection{(Non-)detections of oscillations}
Next, we investigate empirically which parameters are of importance for the detection of oscillations in the data. First, we consider the apparent magnitude distribution of the stars with and without detected oscillations, see Fig.~\ref{mag}. As expected the percentage of stars for which we can detect solar-like oscillations decreases for fainter stars. 

\begin{figure}
\begin{minipage}{\linewidth}
\centering
\includegraphics[width=\linewidth]{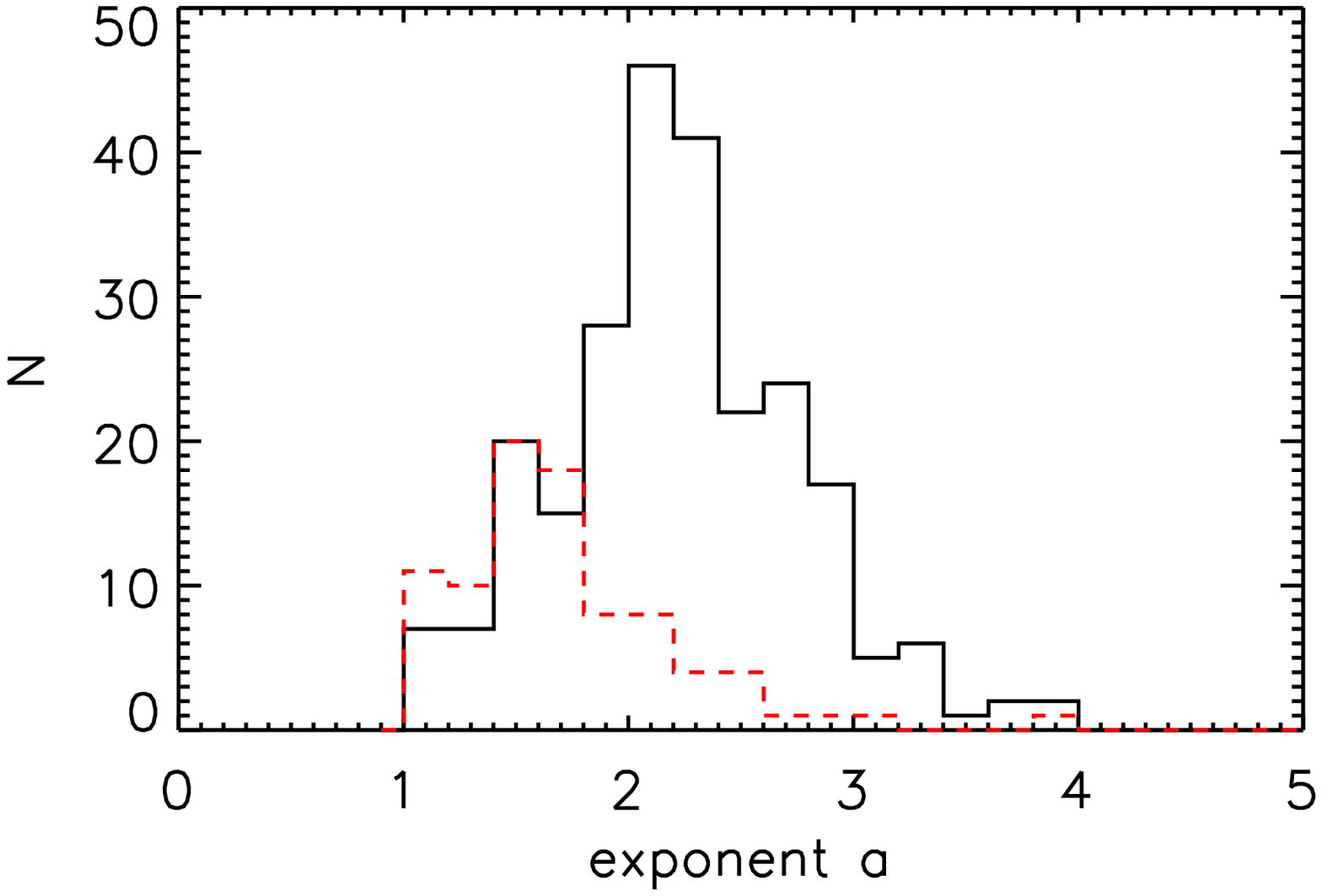}
\end{minipage}
\begin{minipage}{\linewidth}
\centering
\includegraphics[width=\linewidth]{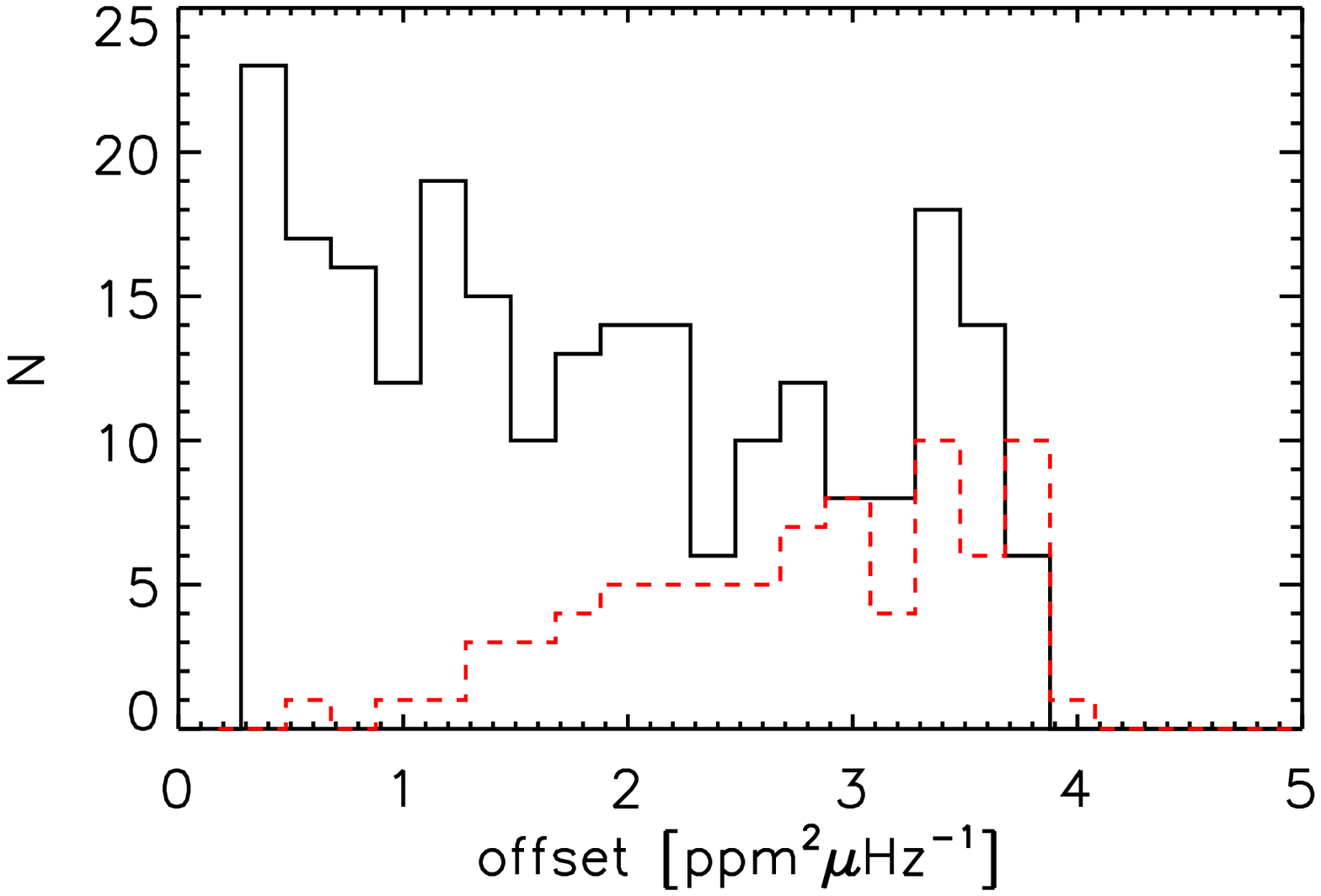}
\end{minipage}
\begin{minipage}{\linewidth}
\centering
\includegraphics[width=\linewidth]{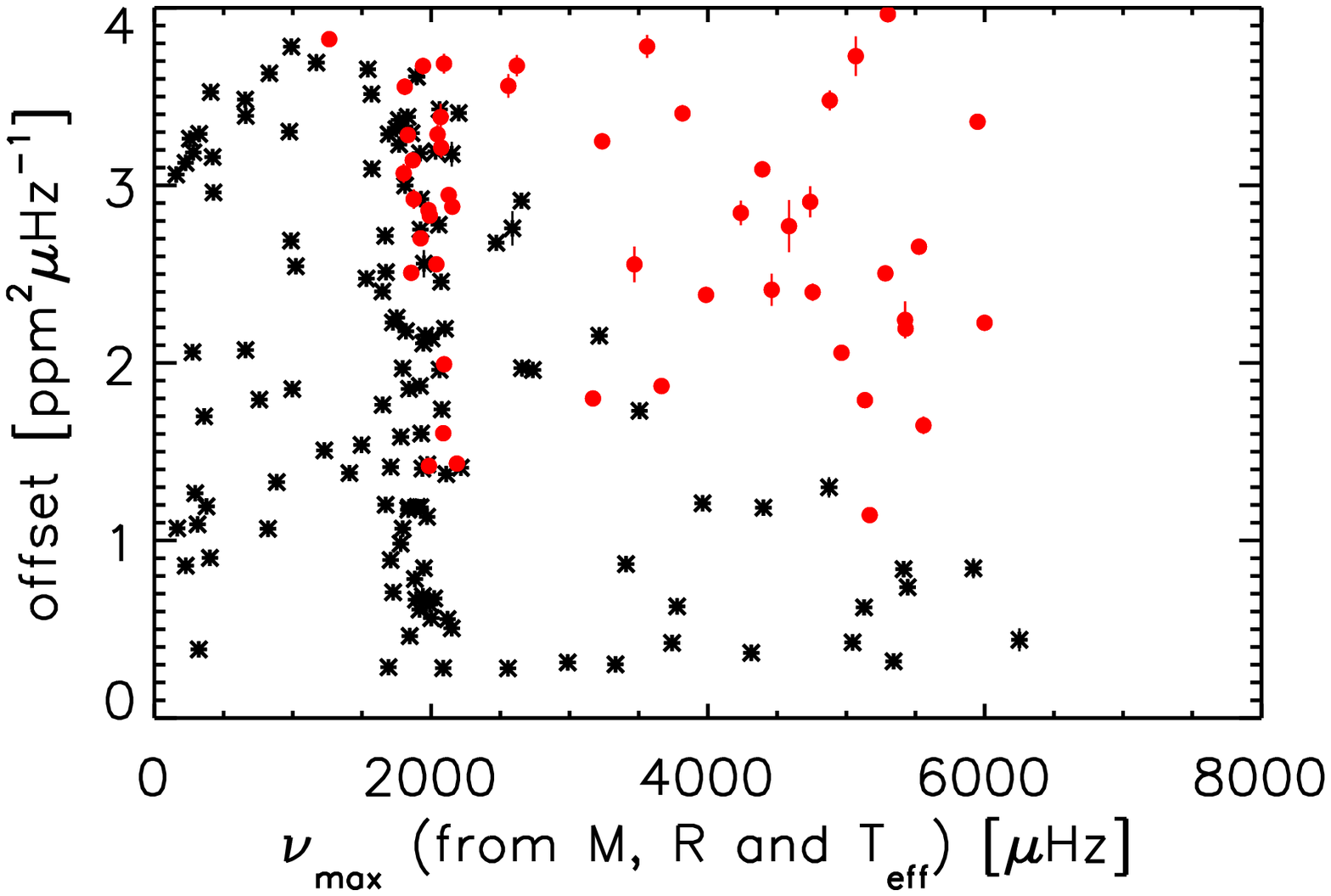}
\end{minipage}
\caption{Distribution of the exponent $a$ (top), the offset (middle) (see Eq.~\ref{backgroundfit}), and offset as a function of input $\nu_{\rm max}$ for stars with detected oscillations (black solid line, black asterisks) and stars without detected oscillations (red dashed line, red dots)}
\label{bg}
\end{figure}

We fitted the background for all stars, independent of whether we did or did not detect any oscillations. From a comparison of the distribution of the fitted parameters, we find that for stars in which we could not detect any oscillations the offset ($b$ in Eq.~\ref{backgroundfit}) is on average larger than for stars in which we could detect oscillations, while the exponent $a$ is typically lower. These distributions are shown in Fig.~\ref{bg}. These results are not unexpected since for stars for which we could not detect oscillations, the offset contains both noise and signal, while for stars with detected oscillations the offset mainly consists of noise. The latter can be seen in the bottom panel of Fig.~\ref{bg}, where we plot the offset as a function of the input $\nu_{\rm max}$. We indeed see that for stars with input $\nu_{\rm max}$ $>$ 3000 $\mu$Hz for which we did not detect oscillations the offset is higher than for stars for which we could detect oscillations in this frequency range. The exponent $a$ influences the slope of the granulation in a log-log plot of a power spectrum. An increase in the offset will decrease the slope and therefore the exponent $a$. From these distributions it might be possible to obtain upper limits on some oscillation parameters. We consider such an investigation beyond the scope of this paper.

\section{Discussion and Conclusions}
With the methods described in Section 3, we could detect solar-like oscillations and their parameters in 260 out of 353 artificial main-sequence stars and subgiants and individual frequencies in 154 stars, not further discussed here. In general, we have tried to be very cautious to reduce the number of false detections. Special care is taken in the identification of the oscillation frequency interval as an incorrectly identified frequency range will imply miss-identifications for all oscillation parameters and the background fitting. Furthermore, parameters such as $\nu_{\rm max}$, $\Delta \nu$ and amplitude per radial mode are determined with two or more (independent) methods. 


The input values for $\nu_{\rm max}$ and $\Delta \nu$ are reproduced by our analyses for the majority of stars. 
Also, for the amplitudes per radial mode, our values are in agreement with the input values.

The widths and thus the life times of the modes can be determined with reasonable accuracy for Kepler stars brighter than 9$^{\rm th}$ magnitude. Our method does not contain detailed fitting, nor does it take into account that lifetimes vary for modes of different degrees. Nevertheless, for the bright stars we do find values consistent with the input relation $\Delta$~$\sim$~$T_{\rm eff}^4$ \citep{chaplin2009a}.

An accurate determination of the value of the background at the oscillation frequencies is important for two reasons. First, the background level is taken into account in the extraction of oscillation parameters such as the amplitude per radial mode, and, secondly, it provides information on the power and time scales of atmospheric parameters. From the fact that we can obtain oscillation parameters which are consistent with the input parameters, we can on the one hand infer that our background estimate in the oscillation frequency range is accurate enough to determine oscillation parameters. On the other hand we still see scatter in the determined time scale and power of the granulation around the input values. 
The scatter in the granulation parameters does not mean per se that the background level in the frequency interval is uncertain, as we fit a function with 6 parameters, but the parameters should be treated with caution when using them for further investigations of activity and granulation.


In terms of our sensitivity to detect oscillations, we found clear evidence that the percentage of stars for which we can detect oscillations decreases for fainter stars. Also, we found evidence that it is harder to detect solar-like oscillations in cooler ($T_{\rm eff}$ $<$ 5500 K) main-sequence stars than in the hotter ($T_{\rm eff}$ $>$ 5500 K) ones. This is because in the simulations (as in real data) the pulsation amplitudes scale with luminosity, with a weaker dependence on temperature.


In conclusion, the analysis tools compiled into a pipeline presented here proved to identify oscillations for a large fraction of artificial main-sequence stars. For the majority of these stars we determine oscillation parameters within 3$\sigma$ of the input values. The existence of such pipelines will be important to be able to perform an asteroseismic analyses on the many stars we expect to become available from Kepler.

\section*{Acknowledgments}
We are grateful to the International Space Science Institute (ISSI) for support provided by a workshop program award. This work has also been supported by the European Helio- and Asteroseismology Network (HELAS), a major international collaboration funded by the European Commission's Sixth Framework Programme. SH, WJC, AMB, YPE, STF, RN acknowledge the support by STFC Science and Technology Facilities Council.  

\appendix
\section{Methodology extension}
Here we provide additional (mathematical) information on the methodology described in Section 3, including determination of errors.\newline
\newline
\textbf{Probability of peaks in PS$\otimes$PS} The probability of the presence of equidistant frequency peaks in the PS$\otimes$PS being due to noise is computed as follows. We compute the probability ($p_1$) that a random variable from a 6 degrees of freedom (dof) $\chi^2$-distribution is larger than the average height of the three peaks at $\Delta \nu$/2, $\Delta \nu$/4 and $\Delta \nu$/6 in the PS$\otimes$PS. Each of these peaks has 2 dof, hence the 6 dof. 

Next, we compute the probability of this occuring by chance at least once over the full $N$ bins of the PS$\otimes$PS. We must also take account of the fact that in practice we oversample the PS$\otimes$PS by a factor of 10, so all bins are not independent. The resulting probability is given
by:
 \[
 P = (1 - p_1)^{\beta N},
 \]
where $\beta$ = 3 has been shown to provide a robust empirical correction for the effect of the oversampling \citep[e.g.,][]{chaplin2002,gabriel2002}. Finally, the probability that the
peaks are \emph{not} due to noise is just $1-P$.


\begin{figure}
\begin{minipage}{\linewidth}
\centering
\includegraphics*[viewport=80 70 420 300,width=\linewidth]{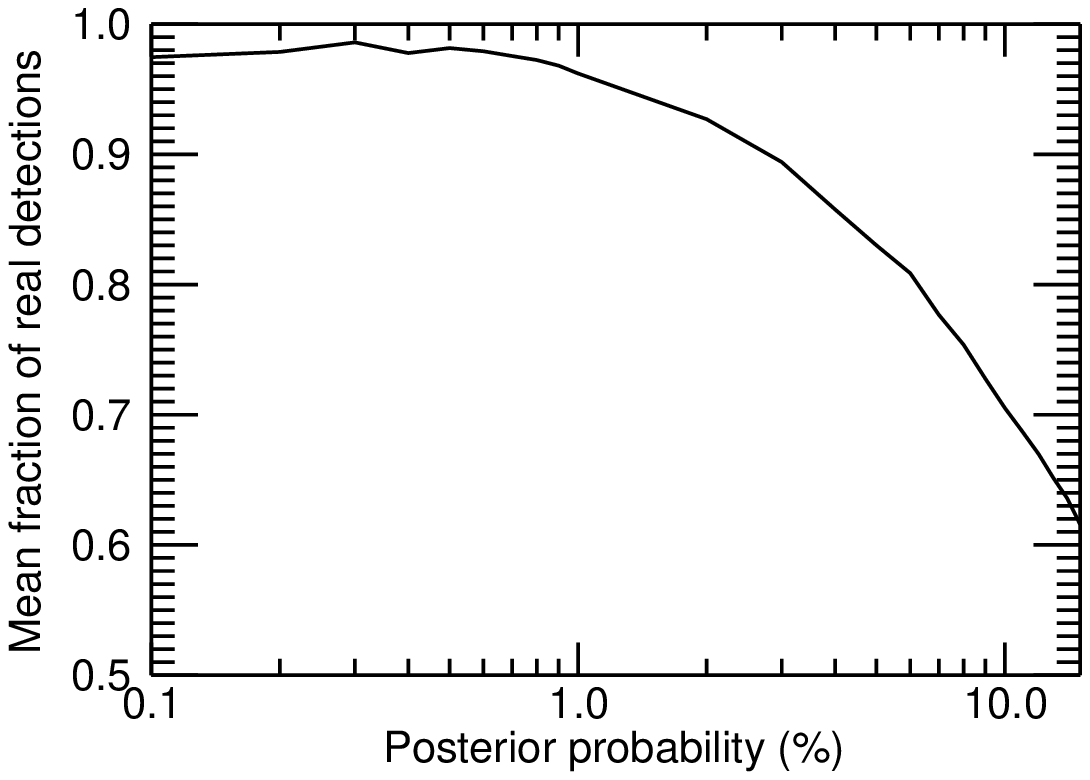}
\end{minipage}
\caption{Fraction of real frequency detections as a function of posterior probability.}
\label{postprob}
\end{figure}

A similar procedure is used to compute the probability of only one peak in the PS$\otimes$PS, such as we use in the computation of the large separation. In these cases $p_1$ is computed as the probability that a random variable from a $\chi^2$ 2-dof distribution is larger than the height of the considered peak in the PS$\otimes$PS.\newline
\newline
\textbf{Gradient of the large spacing} Although results are not discussed in the paper, we can estimate $\delta \Delta \nu$/$\delta$$n$ from the stretching as follows. First, we have
\begin{equation}
\frac{\delta \Delta \nu}{\delta n} = \frac{\delta \Delta \nu}{\delta \nu} \cdot \frac{\delta \nu}{\delta n} = \frac{\delta \Delta \nu}{\delta \nu} \cdot \Delta \nu_0,
\label{1}
\end{equation}
with $\Delta \nu_0$ the average large spacing over the frequency or $n$ range of interest. Now, we may estimate $\delta \Delta \nu$/$\delta \nu$ by differentiating Eq.~\ref{stretch}. To differentiate the second term on the right hand side in Eq.~\ref{stretch} we use the substitution $x$ = $\frac{\nu}{\nu_{\rm c}}$ -1, with $\frac{\delta x}{\delta \nu} = \frac{1}{\nu_{\rm c}}$, which gives:
\begin{equation}
\frac{\delta}{\delta \nu}\left[c\left(\frac{\nu}{\nu_{\rm c}}-1\right)^2\right]  =  \frac{\delta}{\delta \nu}(cx^2)=\frac{2cx}{\nu_{\rm c}} = \frac{2c}{\nu_{\rm c}}\left(\frac{\nu}{\nu_{\rm c}}-1\right).
\end{equation}
So, after including the differential of the first term on the right-hand side, we have
\begin{equation}
\frac{\delta \nu_{\rm stretch}}{\delta \nu} = 1+ \frac{2c}{\nu_{\rm c}}\left(\frac{\nu}{\nu_{\rm c}}-1\right),
\end{equation}
from which we obtain $\Delta \nu$ as a function of frequency:
\begin{equation}
\Delta \nu = \Delta \nu_0 + \frac{2c\Delta \nu_0}{\nu_{\rm c}}\left(\frac{\nu}{\nu_{\rm c}}-1\right).
\end{equation}
Finally, we find that the change in $\Delta \nu$ as a function of $n$ is:
\begin{equation}
\frac{\delta \Delta \nu}{\delta n} = \frac{\delta \Delta \nu}{\delta \nu} \cdot \Delta \nu_0 = \frac{2c\Delta \nu_0}{\nu_c^2} \cdot \Delta \nu_0 = \frac{2c\Delta \nu_0^2}{\nu_c^2}.
\end{equation}
The best-fitting $c$ = $j \cdot s_{\rm max}$ therefore provides a direct estimate of $\delta \Delta \nu$/$\delta n$.\newline
\newline
\textbf{Standard deviation of grouped data} The standard deviation of grouped data is used to compute errors in $\Delta \nu$ computed from the power weighted centroids in the PS$\otimes$PS, where we interpret each feature in the PS$\otimes$PS as compiled of a number of bins with a certain height ($f$) and midpoint ($x$). 
\begin{equation}
s = \sqrt{\frac{\Sigma f x^2 - \frac{(\Sigma fx)^2}{\Sigma f}}{\Sigma f-1}}.
\label{errdnu}
\end{equation}
The same formula is used to compute the error on $\nu_{\rm max}$ from the frequency-binned oscillation power as computed in method I for the amplitudes per radial mode. Here, the total power and central frequency of each bin are interpreted as $f$ and $x$, respectively.\newline
\newline
\newline
\textbf{Bayesian signal hypothesis $\mathbf{H_1}$} For the computation of the probability of observing $x$ if the $H_1$ hypothesis is true, i.e, $p(x|H_1)$ (Eq.~\ref{H1}), we need to integrate over an interval between 0 and $H_s$. To determine $H_s$ we have smoothed the power spectrum over $S$ microHertz. $H_s$ then equals the maximum height of the smoothed spectrum minus the mean background noise level.
To determine the optimum $S$ we performed Monte Carlo simulations and determined the false detection rates. Spectra were generated using the asteroFLAG code \citep{chaplin2008} for 1000 different stars with random inclinations. We then used different values of $S$ to determine $H_s$ and from the obtained candidate frequencies we also determine the ratio of the number of candidates that are actually modes to the total number of mode candidates. The higher this ratio the lower the proportion of false detections.
We also determined the total number of modes that could be detected in each case to ensure that the method was producing a reasonable number of mode candidates.
Note that a mode was counted as a detection if it lay within 2 linewidths of the input frequency and if the posterior probability was less than 0.005. We repeated the simulations for modes with different widths including 0.3, 1.0, 1.7, 2.4, 3.1 and 3.8-times solar widths for oscillations in the range 2000 - 4500 $\mu$Hz. The optimum value of $S$ was found to be $\sim$ 10 $\mu$Hz, even for the smallest line width. 
Similar simulations were performed for stars whose oscillations lay in the range 200 - 450 $\mu$Hz. It was found that for frequencies in this oscillation range it was more appropriate to smooth over a narrower $S$ to determine $H_s$. If the oscillations frequencies are $<$ 450 $\mu$Hz we determined $H_s$ by smoothing over $S$ = 1 $\mu$Hz.


\bibliographystyle{mn2e}
\bibliography{bibasteroflag}

\label{lastpage}

\end{document}